\documentclass[12pt]{article}

\pdfoutput=1

\usepackage{graphics}
\usepackage{amssymb}
\usepackage{amsmath}
\usepackage{hyperref}
\usepackage{bbold}
\usepackage{tikz}
\usepackage{cite}

\usepackage{multicol,multirow}
\usepackage{booktabs, siunitx}
\usepackage{float}

\newcommand{\beq}{\begin{equation}}

\newcommand{\eeq}{\end{equation}}
\newcommand{\bea}{\begin{eqnarray}}
\newcommand{\eea}{\end{eqnarray}}

\begin{document}

\begin{center}
${}$\\
\vspace{100pt}
{ \Large \bf What is the Curvature of
\\ \vspace{10pt} 2D Euclidean Quantum Gravity?
}

\vspace{36pt}

{\sl R.\ Loll}$\,^{\dagger,\star}$ and {\sl T. Niestadt}$\,^{\dagger}$

\vspace{18pt}
{\footnotesize

$^\dagger$~Institute for Mathematics, Astrophysics and Particle Physics, Radboud University \\ 
Heyendaalseweg 135, 6525 AJ Nijmegen, The Netherlands.\\ 

\vspace{5pt}
{\it and}\\
\vspace{5pt}

$^\star$~Perimeter Institute for Theoretical Physics,\\
31 Caroline St N, Waterloo, Ontario N2L 2Y5, Canada.\\
}
\vspace{24pt}

\end{center}


\begin{center}
{\bf Abstract}
\end{center}

\noindent 
We re-examine the nonperturbative curvature properties of two-dimensional Euclidean quantum gravity, obtained as the scaling limit of
a path integral over dynamical triangulations of a two-sphere, which lies in the same universality class as Liouville quantum gravity.
The diffeo\-morphism-invariant observable that allows us to compare the averaged curvature of highly quantum-fluctuating geometries with that of classical
spaces is the so-called curvature profile. A Monte Carlo analysis 
on three geometric ensembles, which are physically equivalent but differ by the inclusion of local degeneracies, 
leads to new insights on the influence of finite-size effects. After eliminating them, we find strong evidence that the curvature profile 
of 2D Euclidean quantum gravity is best matched by that of a classical round four-sphere, rather than the five-sphere found in previous work.
Our analysis suggests the existence of a well-defined quantum Ricci curvature in the scaling limit.

\vspace{12pt}
\noindent

\newpage

\section{Introduction}
\label{sec:intro}

The introduction of a new notion of quantum Ricci curvature suitable for application in nonperturbative quantum gravity \cite{qrc1}
has opened exciting new possibilities for relating the deep quantum regime of Planckian physics to the much more familiar classical
spacetimes described by general relativity, and thus for understanding how the microscopic dynamics of quantum geometry may have left macroscopic
imprints observable in today's universe.

The setting we will consider here is lattice quantum gravity, in its modern form based on dynamical triangulations \cite{Ambjorn2024}, 
a manifestly diffeomorphism-invariant
way to regularize the nonperturbative gravitational path integral. We will only deal with a toy model 
of full quantum gravity (with unphysical dimension and metric signature), which has nevertheless 
garnered much attention in its own right, because of its rich mathematical structure and the multitude of analytical 
and numerical methods that have been used to explore it \cite{Budd2022}.
While many of these solution methods have not been generalized to higher dimensions or Lorentzian signature, both the methodology of lattice gravity in terms of
(causal) dynamical triangulations \cite{review1,review2,ency} and the quantum Ricci curvature \cite{qrc3,curvsumm} 
have been applied successfully beyond the 2D Euclidean case. 

Our main result is based on new Monte Carlo measurements of the curvature properties of 2D Euclidean quantum gravity, more precisely, its so-called curvature profile, 
a diffeomorphism-invariant observable measuring the scale-dependent average quantum Ricci scalar of its quantum geometry. It was
investigated previously in \cite{qrc2}, representing the first application of this new curvature observable in a nonperturbative quantum context. 
Note that on 2D triangulations the quantum Ricci curvature differs from the standard, classical Gaussian curvature, which 
can be expressed in terms of deficit angles at the vertices of the triangulation \cite{regge,curvsumm}. 
Although this is a statement about finite triangulations, rather than their continuum limit, there is no a priori reason why any notion of quantum
curvature should be unique. Classical considerations do not serve as a guide here, since Euclidean quantum gravity
on a two-sphere is purely quantum, with no underlying Einstein equations or a classical limit. 
Instead, it has a fractal geometry and noncanonical scaling laws, characterized by a Hausdorff dimension $d_H\!=\! 4$ and a spectral dimension $d_S\! =\! 2$.
This suggests that any quantum curvature may also scale noncanonically and may not necessarily be associated with a topological
invariant, as is the case for the integrated Gaussian curvature classically.\footnote{The relation between the quantum Ricci curvature and 
the Gauss-Bonnet theorem in two dimensions is discussed further in \cite{Klitgaardthesis}, Sec.\ 3.2.}

Despite the highly nonclassical features of 2D Euclidean quantum gravity,
it was found that the expectation value of its curvature profile can be matched to that of a smooth classical sphere, with a best match to a sphere
of dimension five \cite{qrc2}. This result is doubly surprising, because of the match with a classical manifold at all, and 
because of the appearance of the value 5. 
Although there is currently no theoretical prediction for this number, a dimensionality of five had not been observed previously in the context of 2D quantum gravity. 

However, when interpreting this finding one should keep in mind that the sphere-matching is not straightforward in practice, 
since the curvature profiles for spheres of different dimension are quite similar.
In search of more conclusive evidence to corroborate the result of \cite{qrc2} or otherwise, we have redone the analysis with a broadened scope.
This consisted in performing the measurements on three different ensembles
of regularized geo\-metries, which are known to lead to the same continuum theory. In addition to the regular, simplicial manifolds of
reference \cite{qrc2}, we also used two enlarged ensembles that contain triangulations with local degeneracies, which in the trivalent graph
dual to a triangulation are characterized by the presence of tadpole and/or self-energy graphs. 
Although the associated discrete models are equivalent in the infinite-volume limit, 
the behaviour of observables on \textit{finite} lattices and how well they approximate their continuum values tends to depend on such discretization ambiguities.
This also turns out to be the case for the curvature measurements under investigation. 
Our analysis of these measurements in the enlarged ensembles reveals the presence of certain finite-size effects, which are also present in the
regular ensemble but are masked there by an accidental cancellation with other lattice effects. As a consequence, some of the curve fits made 
previously \cite{qrc2} turn out to be ``too good to be true".   

By carefully investigating both discretization and finite-size effects and selecting data points according to well-motivated prescriptions we are able
to correct these oversights and produce
strong evidence that the curvature profile of 2D Euclidean quantum gravity is best approximated by that of a round four-sphere, with good agreement 
across the different ensembles. This is a satisfying result that brings the dimension of this continuum sphere in line with the Hausdorff dimension 
of this quantum geometry, and hopefully settles the matter for good.

The remainder of the paper is organized as follows. In Sec.\ \ref{sec:qrc}, we discuss quan\-tum-gravitational observables,
recalling the notion of quantum Ricci curvature and why it is physically important. 
Sec.\ \ref{sec:ens} introduces the three different ensembles of 2D triangulations.
Sec.\ \ref{sec:setup} contains a description of the theoretical and numerical set-up of our investigation.
In Sec.\ \ref{sec:res} we present our main numerical results, starting with a series of Hausdorff dimension measurements to test the simulation set-up 
and get a first idea of the influence of the choice of ensemble. This is followed by an investigation of the curvature profile, where we 
critically analyze and extend the methodology and results of \cite{qrc2}, and the role of lattice effects in particular. 
We conclude in Sec.\ \ref{sec:sum} with a summary and outlook. The Appendix contains details of the local geometry of triangulations in the different ensembles.

\section{Observables and quantum Ricci curvature}
\label{sec:qrc}

The curvature properties of spacetime, encoded in the Riemann tensor, are central to how we characterize the physics of classical general relativity. 
The situation is very different in quantum gravity, where the construction of diffeomorphism-invariant observables presents 
major difficulties in any approach (see \cite{Khavkine,Brunetti,Donnelly,Baldazzi} for some recent discussions), 
let alone of observables involving the curvature, whose second-order nature leads to severe ultraviolet (UV) divergences.

Encouraging progress has been made in recent years in constructing concrete diffeomorphism-invariant 
quantum observables in a nonperturbative lattice formulation based on
(causal) dynamical triangulations, aka (C)DT, and in determining their properties analytically (mostly in 2D) or numerically, using Markov chain Monte Carlo methods, see
\cite{review1,review2} for reviews. In this setting, one regularizes the Lorentzian or Euclidean gravitational path integral in $D$ dimensions
in terms of so-called random geometries,
ensembles of piecewise flat, curved spacetimes or spaces, each assembled from a large number of identical\footnote{In the Lorentzian case, there can
be more than one type, because of the different possible orientations of a building block with respect to time and spatial directions.} $D$-dimensional triangular 
building blocks or ``$D$-simplices". 
Following the same logic as in lattice formulations of gauge field theory, one then searches for nontrivial scaling limits as the UV cutoff (the lattice
spacing) is sent to zero. 

To understand the nature of observables in this context, it is important to understand the geometric properties of the configurations that 
are being summed over in the path integral. 
They are nonsmooth, simplicial manifolds, which for vanishing lattice spacing become
nowhere differentiable spaces. They can be thought of as higher-dimensional analogues of the nowhere differentiable paths, which
according to the Wiener measure dominate the path integral of the free nonrelativistic particle \cite{Chaichian2001}. 
For finite lattice spacing, the regu\-larized geometries have well-defined, intrinsic metric properties that allow us to determine distances and volumes. 
In a continuum limit the latter may not scale canonically, i.e.\ not according to their classically expected scaling dimension, reflecting the nonperturbative behaviour
of the quantum geometry. Importantly, the triangulated configurations
are parametrized by a set of discrete data -- geodesic edge lengths and neighbourhood relations -- that describe pure geometry in a nonredundant way,
without the need to introduce local coordinates. Unlike in standard continuum approaches, no gauge-fixing or modding out by the action of the 
diffeomorphism group is therefore required. 

In this manifestly background-independent framework, examples of geometric quantum observables that have been studied in various dimensions are the 
average Hausdorff and spectral dimensions, and in Lorentzian signature the spatial volume of spacetime (aka the Friedmann scale factor) as a function of time.
In the 4D Lorentzian case these observables have produced important evidence for the emergence of a universe with de Sitter-like properties from the nonperturbative 
theory \cite{Ambjorn2007}, and for the presence of a true quantum signature, the dimensional reduction of the spectral dimension at the Planck 
scale \cite{Ambjorn2005}.

In terms of relating Planckian physics to familiar notions of local geometry, what arguably had been lacking from the nonperturbative formulation was a
well-defined notion of local curvature, constructed in terms of the ingredients available 
on the triangulated random lattices. This was accomplished with the introduction of the \textit{quantum Ricci curvature} \cite{qrc1}, which  
provides a generalized notion of Ricci curvature that is also applicable in a strongly quantum-fluctuating regime. It was inspired by the Ollivier-Ricci curvature
on Polish metric spaces (see \cite{ollivier} and references therein). 
Compared to previous constructions based on
finite-difference expressions and deficit angles \`a la Regge calculus \cite{regge}, it has an improved UV behaviour in the continuum limit of the 
4D theory \cite{qrc3,curvsumm}. 

The simplest diffeomorphism-invariant observable one can construct from
the quantum Ricci curvature is the so-called \textit{curvature profile} (essentially its mani\-fold average, see below for a definition).
In 4D quantum gravity this quantity has been shown to match the behaviour of a classical de Sitter universe \cite{qrc3}, in the sense of expectation values, 
further strengthening the relation between fully nonperturbative quantum results and properties of well-known cosmological spacetimes.
The quantum Ricci curvature can also serve as an ingredient in more complex quantum-gravitational observables, 
like diffeomorphism-invariant homogeneity measures of the kind introduced in
\cite{Loll2023}. It has already been used in constructing diffeomorphism-invariant curvature correlators in 2D Lorentzian quantum gravity \cite{correl}. 

\begin{figure}[t]
\centering
\includegraphics[width=0.55\textwidth]{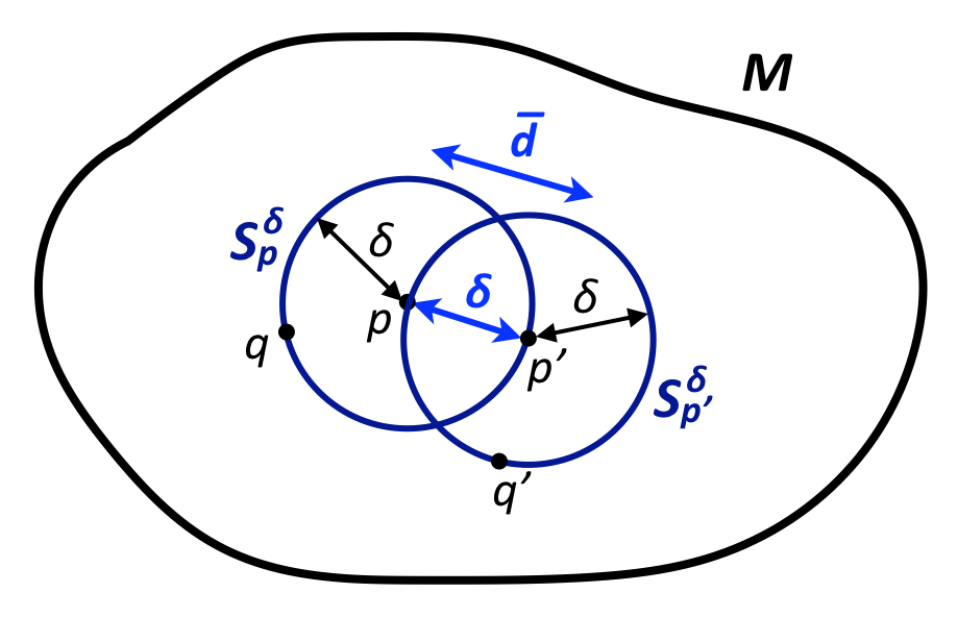}
\caption{The quantum Ricci curvature is determined by comparing the distance $\bar{d}(S_p^\delta,S_{p'}^\delta)$ of two spheres of radius $\delta$ with the 
distance $\delta$ of their centres.}
\label{fig:qrcpic}
\end{figure}

The central idea of the quantum Ricci curvature is to compare the distance between two nearby, small spheres of codimension 1 to the distance between their
centres. It turns out that the ratio of these two distances, as a function of the sphere size and centre distance, captures the local Ricci curvature.
It is easiest to introduce and illustrate this construction on a Riemannian manifold $(M,g_{\mu\nu})$, but the main motivation is its application on the 
nonsmooth, piecewise flat triangulations that appear as configurations in the regularized gravitational path integral, to which we will return below.

For the purposes of the present work, we will focus on the two-dimensional case, where the spheres are one-dimensional circles. 
The geometry of the set-up on a smooth manifold $M$ is depicted in Fig.\ \ref{fig:qrcpic}. 
It consists of two nearby points $p$ and $p'$, separated by a geodesic distance $\delta$, which serve as the centres of two intersecting circles $S^\delta_p$
and $S^\delta_{p'}$, both of radius $\delta$. The associated \textit{average sphere distance} $\bar{d}(S_p^\delta,S_{p'}^\delta)$ is defined by \cite{qrc1}
\begin{equation}
\bar{d}(S_p^{\delta},S_{p'}^{\delta})\! :=\!\frac{1}{\mathrm{vol}\,(S_p^{\delta})}\frac{1}{\mathrm{vol}\,(S_{p'}^{\delta})}
\int_{S_p^{\delta}}\! dq\, \sqrt{h} \int_{S_{p'}^{\delta}}\! dq'\, \sqrt{h'}\ d_g(q,q'),\;\;\; d_g(p,p')\! =\! \delta,
\label{sdist}
\end{equation}
where $d_g(\cdot,\cdot)$ denotes the geodesic distance between pairs of points with respect to the metric $g_{\mu\nu}$, and $h$ and $h'$ are the determinants
of the metrics induced on the two circles. In other words, we perform an average over all distances between point pairs $(q,q')\!\in\! S_p^\delta\!\times\! S_{p'}^\delta$,
and normalize the result by the volumes (lengths) vol($S^\delta$) of the two spheres. 
In terms of the average sphere distance (\ref{sdist}), the quantum Ricci curvature $K_q(p,p')$ at scale $\delta$ is defined by
\begin{equation}
\frac{\bar{d} (S_p^\delta,S^\delta_{p'})}{\delta} =: c_q (1-K_q(p,p')), \;\;\;\;\;\;\; \delta=d_g(p,p'),
\label{ricdefine}
\end{equation}
where $c_q\! =\! \lim_{\delta\rightarrow 0}\bar{d} (S^\delta_p,S^\delta_{p'})/\delta$, which on a two-dimensional Riemannian manifold is given by
$c_q\! \approx \! 1.5746$, and $K_q$ captures the nonconstant remainder. For sufficiently small $\delta$, such that all geodesics between point pairs
exist und are unique, the quotient (\ref{ricdefine}) can be expanded in a power series in $\delta$ at $p$, yielding \cite{qrc2}
\begin{equation}
\frac{\bar{d} (S^\delta_p,S^\delta_{p'})}{\delta}  =1.5746 - 0.1440\, \delta^2\, Ric(v,v) +{\cal O}(\delta^3) ,
\label{expand}
\end{equation}
to lowest nontrivial order, where $Ric(v,v)\! =\! R_{ij} v^i v^j$ denotes the usual Ricci tensor contracted with the unit vector $v$ at $p$ 
pointing to $p'$. Note that in two dimensions $Ric(v,v)$ coincides with the Gaussian curvature at $p$, which in turn is equal to half the Ricci scalar, $R(p)/2$.
The minus signs associated with the curvature terms on the right-hand sides of (\ref{ricdefine}) and (\ref{expand}) reflect the fact that positive (negative) 
Ricci curvature is associated with a decreasing (increasing) slope of $\bar{d}(\delta)/\delta$ as a function of $\delta$, for small $\delta$. For vanishing
curvature, the curvature profile is constant.

The strength of this construction is that it can be used on more general, non\-smooth metric spaces, including the triangulations considered here.
On a two-dimensional triangulation $T$, (integer) distances will be given in terms of the geodesic link distance $d(p,q)$, defined as the number of links 
(edges) in the shortest path along links
between the lattice vertices $p$ and $q$. By definition, the ``sphere"\footnote{Note that this set will in general not literally form a sphere, in the sense of defining a linear
sequence of links (associated with neighbouring vertex pairs) that together form a single circle. For example, it may consist of several disconnected components.} 
$S^\delta_p$ of radius $\delta$ around a vertex $p$ is the set of vertices 
\begin{equation}
S_p^\delta:=\{ q\!\in\! T\, |\, d(q,p)\! =\! \delta\}.
\label{sdef}
\end{equation}
The discrete analogue of the average sphere distance (\ref{sdist}) is given by
\begin{equation}
\bar{d} (S_p^\delta,S_{p'}^\delta)=\frac{1}{N_0(S_p^\delta)} \frac{1}{N_0(S_{p'}^\delta)}\sum_{q\in S_p^\delta} \sum_{q'\in S_{p'}^\delta} d(q,q'),
\;\;\;\;\; d(p,p')\! =\! \delta,
\label{sdistdis}
\end{equation}
where $N_0(S^\delta_p)$ denotes the number of vertices in the sphere $S^\delta_p$. 
A corresponding diffeomorphism-invariant observable, which no longer depends on individual points $p$ and $p'$, is obtained by dividing (\ref{sdist})
by $\delta$ and integrating it over
all (ordered) point pairs $(p,p')$, subject to the condition that they have mutual distance $\delta$. Implementing this directly on a triangulation $T$ 
and normalizing appropriately leads to the so-called \textit{curvature profile} \cite{Brunekreef2020,curvsumm}
\begin{equation}
\frac{\bar{d}_\mathrm{av}(\delta)}{\delta} :=  \frac{1}{{\cal N}_\delta}\; \sum_{p \in T}\sum_{p' \in T} \frac{\bar{d} (S_p^\delta, S_{p'}^\delta)}{\delta} \; \delta_{d(p,p'), \delta} ,
\label{avdis}
\end{equation}
where the Kronecker delta $\delta_{d(p,p'), \delta}$ constrains the distance of $p$ and $p'$ to be equal to $\delta$.
The normalization factor ${\cal N}_\delta$ is given by the double sum 
\begin{equation}
{\cal N}_\delta = \sum_{p \in T}\sum_{p' \in T} \; \delta_{d(p,p'), \delta}
\label{norm1}
\end{equation}
and counts all point pairs $(p,p')$ in $T$ which are a distance $\delta$ apart. 
In analogy with eq.\ (\ref{ricdefine}), we can now extract an \textit{average} quantum Ricci curvature $K_\mathrm{av}(\delta)$, which only depends on the
distance $\delta$, from the curvature profile
(\ref{avdis}) by defining
\begin{equation}
\frac{\bar{d}_\mathrm{av}(\delta)}{\delta} =:  c_\mathrm{av} (1-K_\mathrm{av}(\delta)),
\label{avcurv}
\end{equation}
where on a Riemannian manifold $(M,g_{\mu\nu})$ the constant $c_\mathrm{av}$ depends only on the dimension $D$ and is given by the limit 
$c_\mathrm{av}\! =\! \lim_{\delta\rightarrow 0}\bar{d}_\mathrm{av} (\delta)/\delta$. 

\begin{figure}[t]
\centering
\includegraphics[width=0.55\textwidth]{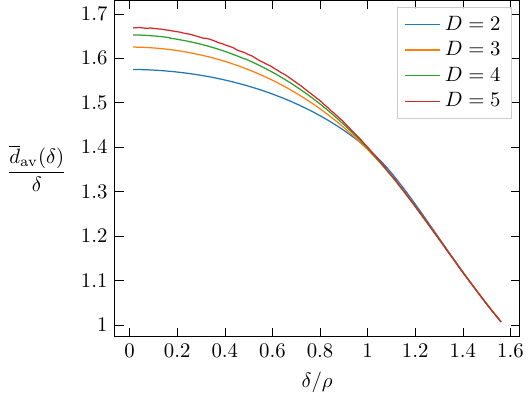}
\caption{Curvature profiles $\bar{d}_\mathrm{av}(\delta)/\delta$ of various $D$-spheres as a function of $\delta$ rescaled by the curvature radius $\rho$. 
}
\label{fig:dspheres}
\end{figure}

Our main objective in the quantum theory will be to measure the
expectation value $\langle \bar{d}_\mathrm{av}(\delta)/ \delta \rangle$ of the curvature profile (\ref{avdis}) in the nonperturbative path integral of 2D Euclidean quantum
gravity in terms of dynamical triangulations, and compare it to curvature profiles of classical manifolds of constant positive curvature, which in previous work were
shown to lead to good matches \cite{qrc2}. The $\delta$-range we are inter\-ested in is a suitable lattice analogue of ``small $\delta$" in the continuum, that is,
$\delta$ is much smaller than the linear size of the entire geometry, but large enough to evade discretization artefacts. 
Generally speaking, maximally symmetric spaces are natural candidates for comparison if one expects that a random-geometric ensemble exhibits approximate symmetry properties
on sufficiently coarse-grained scales.\footnote{It is by no means obvious that such symmetries ``emerge" from quantum gravity formulations based on random geometry
or other microscopic building blocks; this needs to be demonstrated explicitly \cite{Loll2023}.} An additional convenience is the fact that 
one can derive exact integral expressions for the curvature
profiles of maximally symmetric spaces, which can be evaluated numerically \cite{qrc2,Niestadt} and compared with the quantum measurements.

Fig.\ \ref{fig:dspheres} shows
$\bar{d}_\mathrm{av}(\delta)/\delta\!\equiv\! \bar{d}(\delta)/\delta$ for continuum spheres of dimension $D\! =\! $ 2, 3, 4 and 5. Without loss of
generality, the distance $\delta$ is given in units of  $\rho$, the curvature radius of the spheres. In line with our earlier remarks on the nature of the quantum Ricci curvature, 
all curves decrease initially, 
reflecting the presence of positive Ricci curvature. They start at different constant values, which increase with the dimension, but are otherwise qualitatively
similar. Beyond $\delta \!\approx\! \rho$, the curves coincide within the plot's resolution. Before discussing the measurements of the curvature profile in
the quantum theory, we introduce next the different geometric ensembles for which these numerical experiments will be performed.

\section{The three geometric ensembles}
\label{sec:ens}

An essential part of any path integral construction is to specify the space of field configurations the integration should be performed over,
including in the 2D quantum gravity theory considered here.
Since the methodology of obtaining quantum gravity as a suitable scaling limit of a lattice-regularized 
path integral (e.g.\ in terms of DT) applies also in higher dimensions,
it is important to develop an understanding of how the choice of the regularized configuration space influences the 
resulting continuum theory. If lattice formulations of nongravitational quantum field theories and statistical mechanical
models are anything to go by, one expects to find some degree of \textit{universality}, in the sense that
many of the (usually ambiguous) discretization and regularization details will become irrelevant in such a limit, at least when
higher-order phase transitions are involved. 

In full 4D quantum gravity these issues are the subject of ongoing research \cite{renorm,Ambjorn2024,ency}, but fortunately the situation in 2D is much simpler, 
because of the absence of local physical degrees of freedom and UV-divergences, and the analytical control we have over many aspects of the theory.
There are two known universality classes for pure gravity models without matter\footnote{We also do not consider models which include
summing over topologies or allow for topology changes as a function of time.}, 
characterized by a set of critical exponents. Relevant for this paper is the universality class of 2D Euclidean quan\-tum gravity, 
which is characterized by a Hausdorff dimension $d_H\! =\! 4$, spectral dimension $d_S\! =\! 2$ and entropy exponent $\gamma\! =\! -1/2$,
and contains Liouville quantum gravity for $\gamma^L\! =\! \sqrt{8/3}$ \cite{David1988,Distler1988}, 2D Euclidean DT \cite{David1984,Kazakov1985} 
and the Brownian sphere \cite{LeGall2013,Miermont2013}. 
The other universality class is that of Lorentzian or causal models, characterized by $d_H\! =\! 2$, $d_S\! =\! 2$ and $\gamma\! =\! 1/2$ and
contains 2D gravity in proper-time gauge \cite{Nakayama1993}, 2D CDT \cite{Ambjorn1998} and the uniform infinite planar tree \cite{Durhuus2009}. 

The standard choice of configuration space for gravitational Euclidean path integrals in a DT formulation are so-called simplicial manifolds \cite{ACMbook}, also
known as combinatorial triangulations \cite{thor} or combinatorial manifolds \cite{Gallier}, whose definition will be given below. 
The associated \textit{combinatorial ensemble} of such objects will be denoted by ${\cal T}_c$.
Relevant for our investigation is the existence in 2D of two natural generalized ensembles, the \textit{restricted degenerate ensemble}
${\cal T}_r$ and the \textit{maximally degenerate ensemble} ${\cal T}_m$, with ${\cal T}_c\!\subset\! {\cal T}_r\!\subset\! {\cal T}_m$, whose geometric configurations
are allowed to be locally less regular.\footnote{We follow the nomenclature of \cite{thor} for the different ensembles.} 
The most straightforward way to characterize them is in terms of their dual, trivalent planar graphs: ${\cal T}_m$ corresponds to general such graphs,
${\cal T}_r$ to graphs without tadpoles, and ${\cal T}_c$ to graphs without tadpole or self-energy subgraphs. Equivalent characterizations of these ensembles
and their irregularities in terms of the local geometry of the triangulations and the combinatorial data used in the simulations are described in the Appendix.

The three ensembles beautifully illustrate the concept of universality, since they are known to lead to the same
continuum limit of Euclidean 2D quantum gravity.\footnote{How this universality is captured in a matrix model formulation is discussed in \cite{Schneider1998}.} 
Here we will exploit the fact that on finite lattices the different ensembles 
behave differently with regard to finite-size and lattice discretization effects (see e.g.\ \cite{thor} and references therein). However, what represents a best choice
depends on the observable under study and can be difficult to predict beforehand. It turns out that for the cur\-va\-ture profile introduced in Sec.\ \ref{sec:qrc},
we obtain new insights by numerically simulating it on all three ensembles. As we will demonstrate in Sec.\ \ref{sec:curv},
this allows us to draw sharper conclusions on the comparison with continuum spheres and improve on earlier results \cite{qrc2}.

\section{Theoretical and numerical set-up}
\label{sec:setup}

Since the curvature profile of 2D Euclidean quantum gravity has not yet been computed analytically, we will measure it numerically with the
help of Monte Carlo simulations. More precisely, we will measure its expectation value in the two-dimensional gravitational path integral based on DT,
using the three different ensembles ${\cal T}_c$, ${\cal T}_r$ and ${\cal T}_m$ for fixed volumes of up to $N_2\! =\! 240$k, where $N_2(T)$ denotes
the number of triangles in the triangulation $T$. Our starting point is the 2D Euclidean gravitational path integral in the continuum, formally given by
\begin{equation}
Z=\int {\cal D}[g] \, \mathrm{e}^{-S[g]},\;\;\;\;\; S=\Lambda \int_M d^2x\, \sqrt{g}, 
\label{contpi}
\end{equation}
where $\Lambda$ denotes the cosmological constant.
The functional integration is over the space of geometries $[g]$ (metrics modulo diffeomorphisms) on a given manifold $M$, in our case a two-sphere\footnote{The notation $S^{(d)}$ will be used for $d$-spheres as topological
spaces, to distinguish them from the spheres $S^\delta$ of radius $\delta$ defined in eq.\ (\ref{sdef}) as subsets of vertices on a given triangulation.}
$S^{(2)}$. The action $S$ is the Einstein-Hilbert action, which in two dimensions reduces to a cosmological-constant
term, since the integrated Ricci scalar is a topological term which drops out when computing expectation values.

An explicit regularization of (\ref{contpi}) is given in terms of dynamical triangulations, leading to the partition function \cite{Ambjorn1997}
\begin{equation}
Z_\mathrm{DT}(\lambda) =\sum_{T\in{\cal T}} \frac{1}{C(T)}\, \mathrm{e}^{-S_\mathrm{DT}(T)}, \;\;\;\;\; S_\mathrm{DT} (T)=\lambda N_2(T),
\label{discpi}
\end{equation}
where $\lambda\! >\! 0$ is the bare cosmological constant and $\cal T$ is one of the ensembles introduced in Sec.\ \ref{sec:ens}. The weight factor
depends on $C(T)$, the order of the automorphism group of the triangulation $T$, which for large volume $N_2$ is almost always equal to one. 
The continuum limit of the path integral $Z_\mathrm{DT}$ is obtained by fine-tuning the bare cosmological constant $\lambda$ to its critical value $\lambda_c$
from above according to
\begin{equation}
\lambda \rightarrow \lambda_c +a^2 \Lambda+ O(a^3),
\label{lambda}
\end{equation}
where $a$, the edge length of the triangular building blocks, plays the role of a UV cutoff that is taken to zero in this limit. The limit can be computed analytically
using e.g.\ combinatorial or matrix model methods \cite{David1984,Kazakov1985,Ambjorn1997}. For our purposes, we will use a numerical implementation of 
the path integral (\ref{discpi}). In this setting, the expectation value of a diffeomorphism-invariant geometric observable ${\cal O}(T)$ is given by 
\begin{equation}
\langle {\cal O}\rangle =\frac{1}{Z_\mathrm{DT}}\sum_T \frac{1}{C(T)}\, {\cal O}(T)\, \mathrm{e}^{-S_\mathrm{DT}(T)}.
\label{expval}
\end{equation}
Our actual computations will take place in fixed-volume ensembles, as customary, with corresponding expectation values 
\begin{equation}
\langle {\cal O}\rangle_{N_2} =\frac{1}{Z_{N_2}}   \sum_{T|_{N_2}} \frac{1}{C(T)}\, {\cal O}(T),\;\;\;\;\;\; Z_{N_2}=  \sum_{T|_{N_2}} \frac{1}{C(T)},
\label{expfix}
\end{equation}
where the sums are over triangulations of a given discrete volume $N_2$.
The fixed-volume partition function $Z_{N_2}$ in (\ref{expval}) is related to the DT path integral (\ref{discpi}) for cosmological constant $\lambda$ 
by a Laplace transform,
\begin{equation}
Z_\mathrm{DT}(\lambda) =\sum_{N_2} \mathrm{e}^{-\lambda N_2} Z_{N_2}.
\label{}
\end{equation}
We approximate an expectation value (\ref{expfix}) by averaging its measured value over a 
sequence $T_1$, $T_2$, $T_3$, ..., $T_n$ of independent configurations, generated by a Markov chain Monte Carlo (MCMC) algorithm (see \cite{BuddMC} for
an introduction). Its behaviour in the infinite-volume limit is extrapolated from measurements at various fixed finite volumes 
$N_2$ by finite-size scaling, a standard tool for analyzing statistical systems (see e.g.\ \cite{NB}). 

\begin{figure}[t]
\centering
\includegraphics[width=0.7\textwidth]{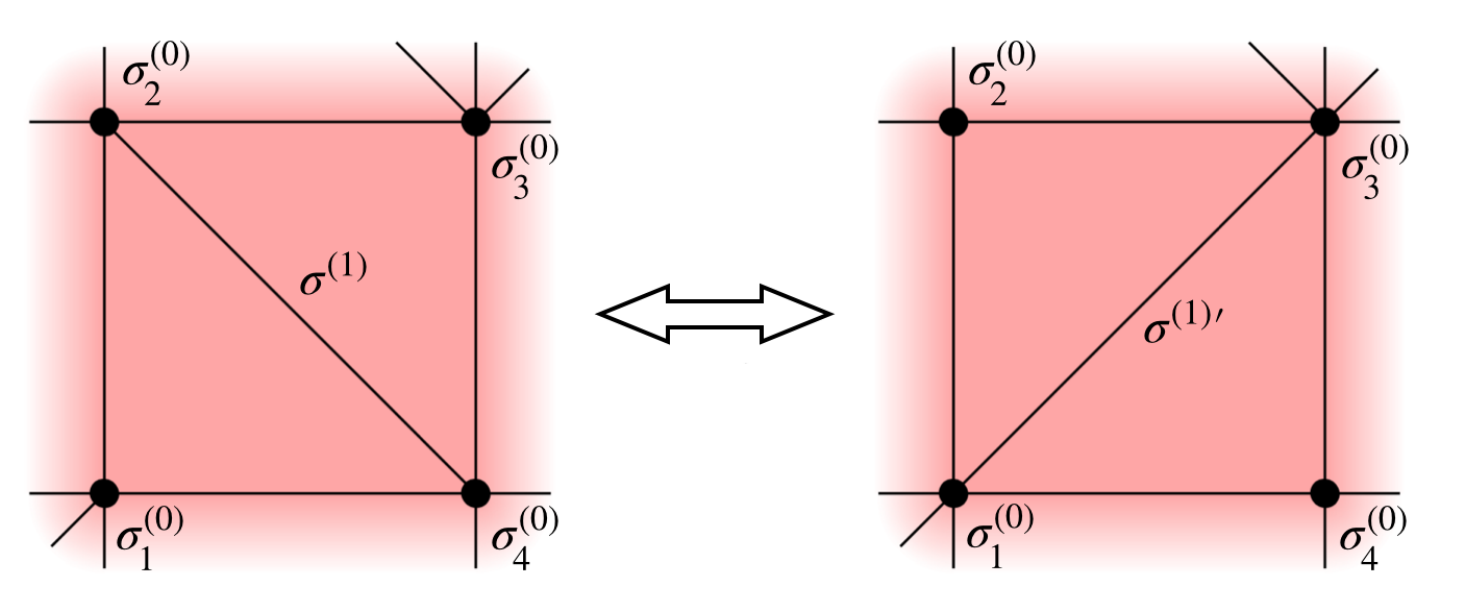}
\caption{Flip move and its inverse: the edge $\sigma^{(1)}$ between the vertices $\sigma^{(0)}_2$ and $\sigma^{(0)}_4$ is replaced by the edge $\sigma^{(1)}{}'$
between the vertices $\sigma^{(0)}_1$ and $\sigma^{(0)}_3$, and vice versa.
}
\label{fig:flip}
\end{figure}

The local Monte Carlo move we use for updating
geometries is the flip move. It consists of choosing an edge $\sigma^{(1)}$ in $T$ and substituting the two triangles sharing this edge by another
pair of triangles as shown in Fig.\ \ref{fig:flip},
which in the planar representation of the figure amounts to flipping the diagonal $\sigma^{(1)}$ of the ``square" formed by the triangles to the new diagonal edge $\sigma^{(1)}{}'$. 
The flip move is known to be ergodic in the combinatorial ensemble, which means that from any configuration $T\!\in\! {\cal T}_c$ one can reach any other 
$T'\!\in\! {\cal T}_c$ by a finite number of flip moves. The ergodicity proof can be extended to the two degenerate ensembles by showing that any local
degeneracy can be removed by performing a finite number of flip moves \cite{Niestadt}. During an update in a given ensemble $\cal T$, an edge of a triangulation 
$T\!\in\! {\cal T}$ is selected uniformly at random as a candidate for a flip move. The move is accepted if the resulting triangulation $T'$ lies again in $\cal T$, 
and rejected otherwise. The approximate rejection rates for the ensembles ${\cal T}_c$, ${\cal T}_r$ and ${\cal T}_m$ were 1/4, 1/6 and 0 respectively.
After a careful assessment of autocorrelation times, we set the sweep size between subsequent measurements of both the average Hausdorff dimension 
(in Sec.\ \ref{sec:haus}) and 
the average sphere distance (in Sec.\ \ref{sec:curv}) to 50, to ensure the statistical independence of subsequent samples.\footnote{A sweep is defined here 
as $N_2$ attempted moves.}
Readers interested in further technical details of the Monte Carlo simulations for this and other systems of dynamical triangulations may 
consult \cite{Niestadt,Ambjorn1997,simul} and references therein. The simulation code for the results presented below can be found in an online repository \cite{code}.

\section{Measurements and results}
\label{sec:res}
\subsection{Hausdorff dimension}
\label{sec:haus}

We started by investigating the influence of the ensemble on the Hausdorff dimension, which at the same time served
as a cross-check of our code and a point of comparison with previous numerical studies of 2D Euclidean quantum gravity.
Like for the quantum Ricci curvature, its measurement involves one-dimensional spheres, and can therefore give us a first idea of the
role played by the chosen ensemble. The (local) Hausdorff dimension $d_H$
can be defined as the leading power in the behaviour of the expectation value of the average sphere volume as a function of the sphere radius $r$,
\begin{equation}
\langle \overline{\mathrm{vol}(S^r)}\rangle_{N_2} \sim r^{d_H-1}
\label{svol}
\end{equation}
for small $r$. 
We noted earlier that the point set $S^r_p$ at a vertex $p\!\in\! T$ defined by eq.\ (\ref{sdef}) in general does not form a true sphere,
in the sense that the edges between pairs of neighbouring vertices do not line up to form a $S^{(1)}$
topologically. For $r\! =\! 1$, examples of this for $S^1_p$, with $p\! =\! {\sigma^{(0)}_2}$, are given in Figs.\ \ref{fig:flathat} and \ref{fig:maxdegflat} of the appendix. Similar 
phenomena are present for larger $r$, also in the combinatorial ensemble. The overbar in eq.\ (\ref{svol}) denotes a configuration average, 
which for a local\footnote{Here and elsewhere, ``(quasi-)local" should be understood in a lattice sense; due to the discretization, a (quasi-)local quantity
associated with a vertex $p$ will typically depend on fields not just at $p$ but also at (sub-)simplices in a small neighbourhood of $p$.} geometric
quantity ${\cal O}(p)$ on a triangulation $T$ is defined by
\begin{equation}
\overline{\cal O}|_T:=\frac{1}{N_0}\sum_{p\in T}{\cal O}(p),
\label{oav}
\end{equation}
where $N_0$ is the number of vertices in $T$. Any local scalar must be summed or averaged over all of $T$ to obtain a diffeomorphism-invariant observable,
whose expectation value can then be computed in the given ensemble $\cal T$.

\begin{figure}[t]
\centering
\includegraphics[width=1.01\textwidth]{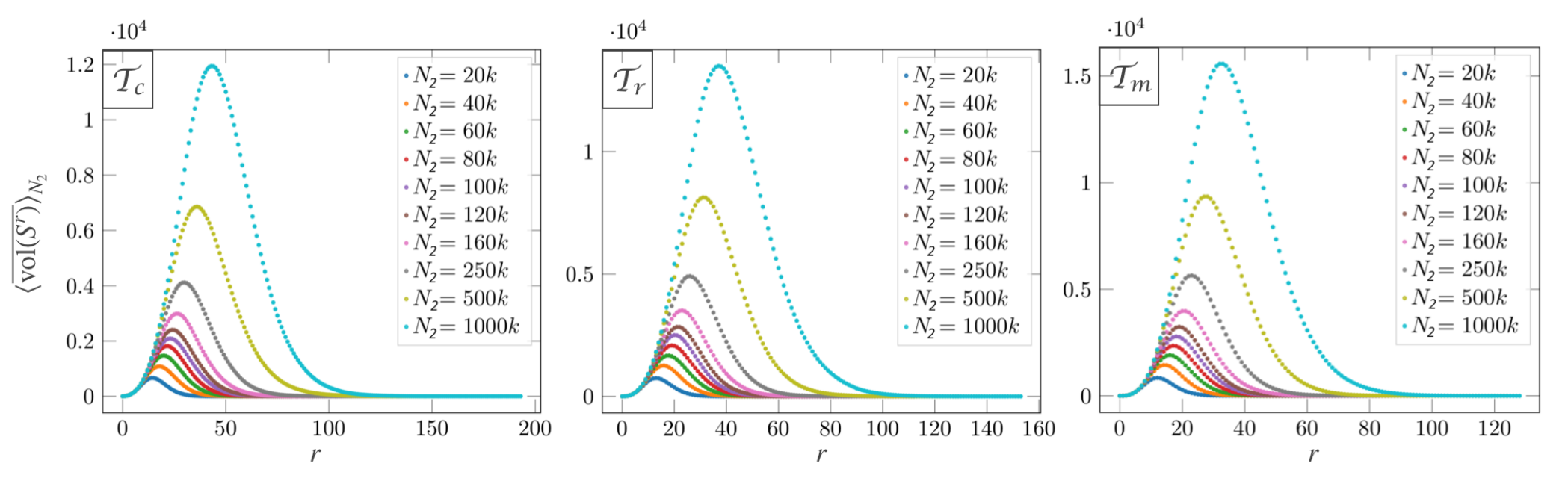}
\caption{Measured expectation values $\langle \overline{\mathrm{vol}(S^r)}\rangle_{N_2}$ of the average volumes of one-dimensional spheres $S^r$
of radius $r$ in the three ensembles ${\cal T}_c$ (left), ${\cal T}_r$ (middle) and ${\cal T}_m$ (right), for volumes $N_2\in [20k,1000k]$. 
}
\label{fig:distr3E}
\end{figure}

A variety of methods to determine the average Hausdorff dimension has been employed in previous work. Analytical considerations \cite{dim2dnum} 
suggest the presence of
a universal function $\rho(x)$ in the quantum-gravitational continuum theory, which is related to the expectation value of sphere volumes for all (not just small) radii. 
In a lattice implementation this leads to the scaling ansatz 
\begin{equation}
\lim_{N_2\rightarrow\infty} N_2^{1/d_H} \rho_{N_2}(N_2^{1/d_H} x)=\rho(x)
\label{rhoscale}
\end{equation} 
in the infinite-volume limit, where $x\! :=\! N_2^{-1/d_H} r$, and $\rho_{N_2}(r)$ is a normalized ver\-sion of the expectation value,
\begin{equation}
\rho_{N_2}(r):=\frac{ \langle \overline{\mathrm{vol}(S^r)}\rangle_{N_2}}{N_0}=\frac{ \langle \overline{\mathrm{vol}(S^r)}\rangle_{N_2}}{2+N_2/2}
\approx 2\ \frac{ \langle \overline{\mathrm{vol}(S^r)}\rangle_{N_2}}{N_2},
\label{rhodef}
\end{equation}
where we have used an identity valid for triangulated two-spheres and in the last step dropped a subleading term. 
The same Hausdorff dimension $d_H$ as before appears in the global scaling relation (\ref{rhoscale}), reflecting the fact that in pure Euclidean 2D quantum gravity 
the local notion of Hausdorff dimension defined by (\ref{svol}) extends to a global one.
Note that for any triangulation $T$ of $S^{(2)}$ with $N_0$ vertices and for an arbitrary vertex $p\!\in\! T$, we have
\begin{equation}
\sum_{r\geq 0} \mathrm{vol}(S^r_p) =N_0,
\label{norm2}
\end{equation}
since every vertex of $T$ will be contained in exactly one sphere $S^r_p$. This implies $\sum_{r} \rho_{N_2}(r)\! =\!1$, justifying the normalization
in the definition of (\ref{rhodef}).
The unnormalized expectation values $\langle \overline{\mathrm{vol}(S^r)}\rangle_{N_2}$ of the average sphere volumes as a function of the radius $r$ 
in the three ensembles are depicted in Fig.\ \ref{fig:distr3E}. They have a slightly skew-symmetric shape around their maximum, 
which is familiar from previous investigations. 
This is not a finite-size effect, but a known property of two-dimensional Euclidean quantum gravity \cite{dim2dnum}.
The main distinguishing feature between the three ensembles is that
for identical two-volume, a more degenerate ensemble correlates with a sharper peak, whose location $r_0$ is at a smaller radius.

We will not describe the technical details of the various Hausdorff dimension measurements, which was done by previous authors and is not
relevant for what follows, and only summarize our results. 
All methods of extracting the Hausdorff dimension described in this subsection are based on $10^4$ measurements each
for ten different two-volumes in the range $N_2\!\in\! [20\mathrm{k},1000\mathrm{k}]$, for each of the three ensembles.

The approach of \cite{Catterall1995} was to extract esti\-mates for $d_H$ from the location and height of
the peak of the distribution $\rho_{N_2}(r)$ as a function of volume, which from relation (\ref{rhoscale}) are expected to scale like
\begin{equation}
r_0\propto N_2^{1/d_H}\;\;\; \mathrm{and} \;\;\; \rho_{N_2}(r_0)\propto N_2^{-1/d_H} 
\label{scalings}
\end{equation}
respectively. The peak location is determined by fitting the distribution to 
the functional form
\begin{equation}
\rho_{N_2}(r)=P_4(r)\exp(-\alpha r^\beta),
\label{poly}
\end{equation}
for a fourth-order\footnote{According to \cite{Catterall1995}, $l\! =\! 4$ is the lowest order of a polynomial $P_l$ that leads to a ``reasonably good" fit with
the ansatz (\ref{poly}).} polynomial $P_4(r)$ with associated fitting coefficients and two additional fitting parameters $\alpha$ and $\beta$. The maximum
is then found by using the fitted function, and the locations and heights of the maxima for different $N_2$ can subsequently be fitted to (\ref{scalings}). 
The estimates for $d_H$ we obtained from this procedure are listed in the first two data columns of Table \ref{haustable}. We observe that for both methods 
the Hausdorff dimensions measured in the combinatorial and restricted degenerate ensembles 
significantly underestimate the analytically known value of $d_H\! =\! 4$. This is a familiar feature from the long history of measurements of this observable
in 2D EDT (see e.g.\ the discussion in \cite{Catterall1995}). 
Our measurements show a clear improvement when using a less regular lattice, presumably because lattices that allow for a smaller
coordination number (number $n$ of edges meeting at a vertex) explore the local geometry more effectively and reduce finite-size effects.\footnote{The minimal 
coordination number is $n_\mathrm{min}\! =\! 3$, 2 and 1 in the ensembles ${\cal T}_c$, ${\cal T}_r$ and ${\cal T}_m$ respectively.} The authors of \cite{Catterall1995}
worked with the maximally degenerate ensemble ${\cal T}_m$, took data for six different lattice sizes in the range $N_2\!\in\! [1\mathrm{k},32\mathrm{k}]$,
and reported $d_H\! =\!\num{3.835 \pm 0.059}$ from peak location and $d_H\! =\!\num{4.040 \pm 0.098}$ from peak height measurements. 
Comparing these values with our results, they lie just outside each others' statistical error bars, which we take as an indication that they are broadly compatible.

\begin{table}[t]
\centering
\begin{tabular}{ |c|c|c|c|c|p{2.3cm}|p{2.3cm}|p{2.3cm}| p{2.3cm}| }
\hline
&\multicolumn{4}{|c|}{$d_H$} \\
\hline
ensemble& peak location & peak height & best overlap & best collapse \\
\hline
${\cal T}_c$ & \num{3.625 \pm 0.011} & \num{3.677 \pm 0.018} & ---  & \num{3.880 \pm 0.010} \\
${\cal T}_r$  & \num{3.767 \pm 0.014} & \num{3.869 \pm 0.018} & \num{3.917 \pm 0.095}  & \num{3.798 \pm 0.011} \\
${\cal T}_m$  & \num{3.930 \pm 0.015} & \num{3.922 \pm 0.017} & \num{3.930 \pm 0.050} & \num{3.914 \pm 0.013} \\
\hline
\end{tabular}
\caption{Estimates for the Hausdorff dimension $d_H$ in the three ensembles, using a variety of methods as explained in the text.}
\label{haustable}
\end{table}

A more refined method to extract the Hausdorff dimension, which can also give a clearer indication of the presence of finite-size scaling, is to 
compare the entire curves of rescaled distributions $\rho_{N_2}$, and find the $d_H$ that leads to a best overlap of the curves. 
Following \cite{dim2dnum}, we considered the distributions as functions of the rescaled radial variable 
\begin{equation}
\tilde{x} = \frac{r+a}{N_2^{1/d_H}+b}, 
\label{resc}
\end{equation}
with additional phenomenological shift parameters $a$ and $b$. We could not get this fitting procedure to work for the combinatorial ensemble, where
the quality and quantity of our data was apparently insufficient, but there was no such problem for the degenerate ensembles. 
The column labelled ``best overlap" in Table \ref{haustable} contains the corresponding Hausdorff
dimensions we extracted from measuring for volumes of up to 1000k. Both values are closer to 4, but at the expense of signi\-ficantly larger error bars. 
This is compatible with the finding $d_H\! =\! \num{3.9 \pm 0.2}$ of \cite{dim2dnum}, obtained on the largest ensemble ${\cal T}_m$ for a variety of system sizes 
$N_2\!\in\! [1\mathrm{k},32\mathrm{k}]$. Note that the latter implementation worked with an enlarged set of Monte Carlo moves,
which in addition to the flip move included a nonlocal surgery move. 

Lastly, we used a variant of the previous method, which was introduced in \cite{barkley}, where again entire curves are brought to overlap. 
The difference is that one looks for an optimal collapse to a specific reference
curve, given by the distribution $\rho_{N_{2}}(x)$ for the maximal volume, in our case $N_{2,\mathrm{max}}\! =\! 1000$k. 
Also this method uses a constant shift $s$ of the argument $x$ for a given ensemble $\cal T$ to obtain a best curve collapse. 
The quality of the collapses we found with this method\footnote{Due to 
insufficient statistics, we had to omit a subleading term in $N_2/N_{2,\mathrm{max}}$ used in \cite{barkley}.} 
were very good, especially for the degenerate ensembles 
(see \cite{Niestadt} for further details), and resulted in the Hausdorff dimension estimates collected in the last column of Table \ref{haustable}.
They follow the pattern of the other columns in that the maximally degenerate ensemble gives the dimension closest to the correct value of 4.\footnote{We
do not have a clear explanation for why the dimensions in the last column do not grow monotonically; a closer examination of this issue lies beyond the scope of this paper.}
The sizes of the statistical error bars are smaller than those for the other methods, but the actual estimates of $d_H$ still do not lie within a standard deviation
of $d_H\! =\! 4$, but are systematically lower, presumably due to finite-size effects.
The work of \cite{barkley}, which used quadrangulations instead of triangulations, has shown that a dedicated numerical effort   
and much larger lattices (of up to $2^{24}\!\approx\! 17$m squares) are needed to do significantly better, in that case
yielding a ``precision estimate" of $d_H\! =\! \num{3.9970\pm 0.0013}$.

We conclude that for the lattice sizes and statistics used by us, the choice of ensemble has a significant influence on the measurements of the Hausdorff dimension,
as illustrated by Table \ref{haustable}. The results for this particular observable are consistently closer to the known continuum value when we use
the maximally degenerate ensemble.
At the same time, we found that our results are compatible with those of previous authors, wherever a comparison was possible, 
providing a consistency check for our code and numerical set-up.

\subsection{Curvature profile}
\label{sec:curv}

Our investigation of the diffeomorphism-invariant 
curvature profile $\bar{d}_\mathrm{av}(\delta)/\delta$ of eqs.\ (\ref{avdis}) and (\ref{avcurv}) in the quantum theory proceeds in several steps. 
We begin in Sec.\ \ref{sub1} by repeating the analysis of reference \cite{qrc2} for the combinatorial ensemble, 
whose results we are trying to corroborate, and then extend the procedure to 
the two degenerate ensembles. Although this at first glance seems to confirm the conclusion that $D\! =\! 5$ is the best fitting dimension \cite{qrc2}, 
the lower quality of the fits for the ensembles ${\cal T}_r$ and ${\cal T}_m$ motivates a closer look at the choice of the interval $\delta\!\in\! [\delta_\mathrm{min},
\delta_\mathrm{max}]$, for which the measurements are considered reliable, i.e.\ are not significantly affected by discretization or finite-size effects. 
In Sec.\ \ref{sub2}, we introduce a new method to determine the lower bound $\delta_\mathrm{min}$ for a given ensemble and lattice size. 
This brings the quality of the fits for the different ensembles into line with each other and reveals the presence of significant finite-size effects 
for data taken for small lattice volumes, including for the combinatorial ensemble, which had previously been masked. 
In Sec.\ \ref{sub3}, we address these effects by introducing a criterion based on the behaviour of the average sphere volumes of Sec.\ \ref{sec:haus}  
to determine individual thresholds $\delta_\mathrm{max}$ per volume and ensemble above which data points should be discarded.
The overall effect of this refined analysis are fits of very good quality and with small error bars, which across the different ensembles are in agreement with
each other and strongly favour $D\! =\! 4$ as the best fitting dimension for the curvature, in contrast with the earlier finding of \cite{qrc2}.

\subsubsection{Analysis of the combinatorial ensemble reloaded}
\label{sub1}

We proceed along the lines of reference \cite{qrc2}, which for completeness will be recalled here. For each sample triangulation $T$ in the Monte Carlo
algorithm described in Sec.\ \ref{sec:setup}, we pick a vertex $p\!\in\! T$ at random, which is kept fixed during the following steps. 
Sequentially for each lattice distance $\delta\! = \! 1,\, 2,\dots,\, 15$,
we pick another vertex $p'$ uniformly at random from the vertex set making up the sphere $S^\delta_p$ and determine the average sphere distance 
$\bar{d} (S_p^\delta,S_{p'}^\delta)$ between the two spheres of radius $\delta$ around $p$ and $p'$ according eq.\ (\ref{sdistdis}). 
This set of measurements is then repeated for $2\!\times\! 10^4$ triangulations for each of the sizes $N_2\! =\! 20$, 30, 40, 60, 80, 120, 160 and 240k.
We thus perform fewer measurements at $20$k and $30$k than in \cite{qrc2}, but in our view this leads to sufficiently
small statistical errors. The only other methodological difference with \cite{qrc2} is that we also collect data for the two degenerate ensembles.  

The above prescription contains some shortcuts to save computation time. Since generating a new triangulation $T$ is much less time-consuming
than performing the manifold averages in the definition (\ref{avdis}) of the curvature profile, one substitutes the manifold average by the 
ensemble average, which is a well-motivated and standard assumption for measurements of a quasi-local quantity on sufficiently large volumes.
Moreover, for the measurements on a given configuration $T$, the point $p$ in the point pairs $(p,p')$ is kept fixed as the distance $\delta$ is varied,
which allows the construction of $S^{\delta +1}_p$ from $S^\delta_p$. Lastly, it turns out that the sampling of point pairs $(p,p')$
at distance $\delta$ just described is not strictly uniform in the space of all such pairs in a given triangulation $T$, but has a dependence on the ratio
between vol$(S^\delta_p)$ and its configuration average $\overline{\mathrm{vol}(S^\delta)}|_T$. This was pointed out and investigated in the
context of 2D CDT \cite{correl}, where the dependence was found to be relatively mild and primarily affecting small distances $\delta$. 
The same observation holds also in our context. We will nevertheless stick to the nonuniform sampling, since it will allow us to compare our results directly
with those of \cite{qrc2}, and since there is no indication that it will make a difference in the continuum limit. 

One feature that remains to be addressed before turning to the measurements is how to identify the lattice analogue of the $\delta$-independent
constant $c_\mathrm{av}$ on the right-hand side of eq.\ (\ref{avcurv}). This is important when comparing to the curvature profile of continuum spaces,
in our case round spheres of dimension $D$ and curvature radius $\rho$,
because the value of this constant is nonuniversal and depends on the discretization \cite{qrc1}.
Due to short-distance lattice artefacts, defining $c_\mathrm{av}$ by a limit $\delta\!\rightarrow\! 0$, as was done in the smooth classical case, is not a meaningful prescription. 
It should instead be chosen at a value $\delta_\mathrm{min}$ as close as possible to $\delta\! =\! 0$, but outside the small-$\delta$ region dominated by 
discretization artefacts, defining $c_\mathrm{av}\!:=\! \langle\bar{d}_\mathrm{av}(\delta)/\delta\rangle |_{\delta=\delta_\mathrm{min}}$.
The optimal choice of $\delta_\mathrm{min}$ and how different choices affect the final results should be examined on a case-by-case basis. 
For two-dimensional triangulations studied so far, including regular lattices and Delaunay triangulations \cite{qrc1}, and configurations of DT \cite{qrc2} and CDT quantum gravity \cite{Brunekreef2021,correl}, significant lattice artefacts have been found in the region $\delta \lesssim \delta_\mathrm{min} \!\approx\! 5$.   
As we will see in Sec.\ \ref{sub2} below, the different ensembles used in our re-examination of the curvature profiles also require a fresh look at $\delta_\mathrm{min}$.

\begin{figure}[t]
\centering
\includegraphics[width=0.65\textwidth]{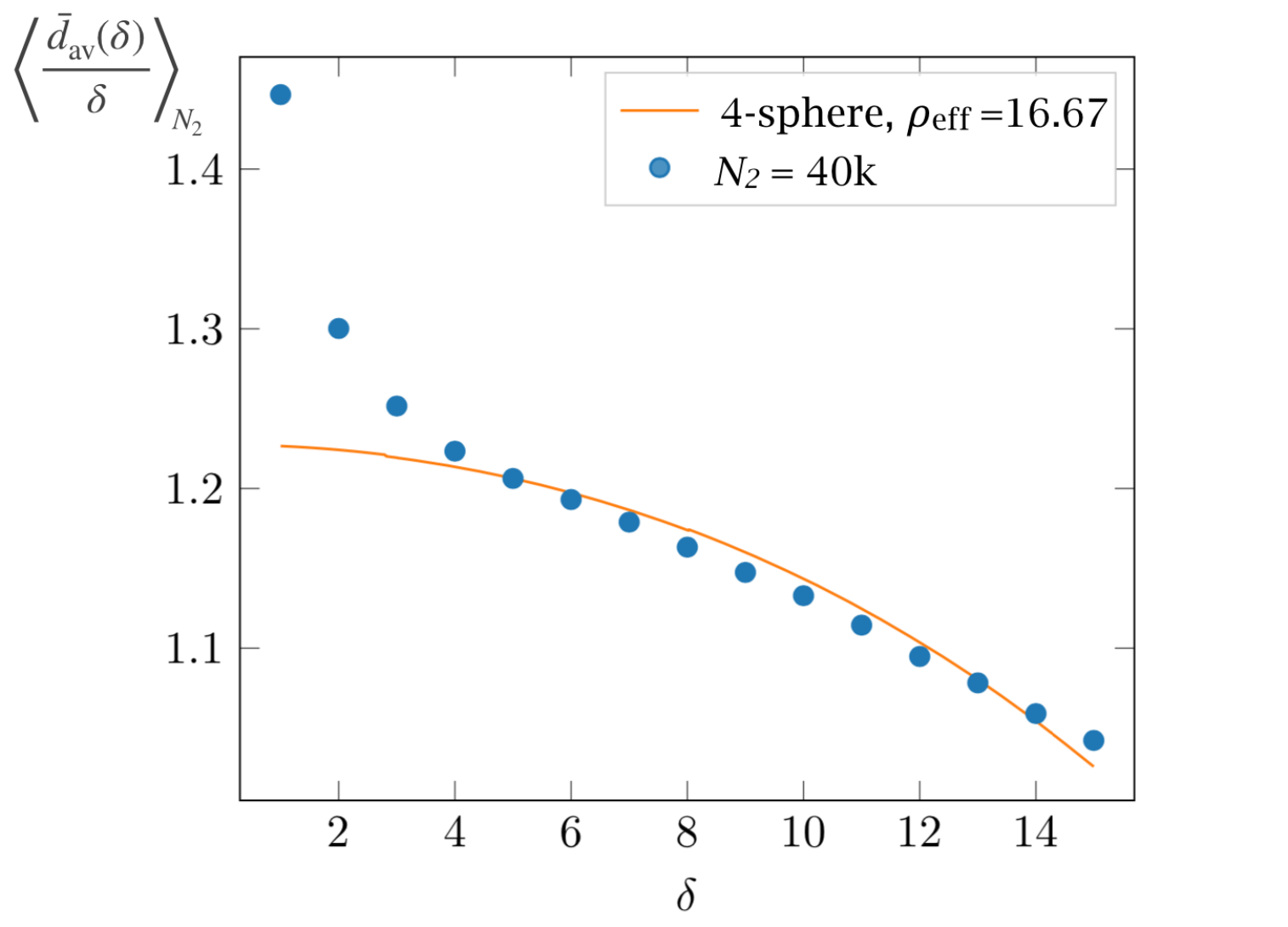}
\caption{Curvature profile $\langle\bar{d}_\mathrm{av}(\delta)/\delta\rangle_{N_2}$ as a function of the distance $\delta$ 
for combi\-na\-torial triangulations of volume $N_2\! =\! 40$k (blue dots, error bars 
are smaller than dot size),
with best fit to the curvature profile of a continuum four-sphere with curvature radius $\rho_\mathrm{eff}\! =\! 16.67$
(yellow curve).
}
\label{fig:CPcomb40k}
\end{figure}

To begin with, we discuss
our findings for the combinatorial ensemble ${\cal T}_c$, where for the time being we follow \cite{qrc2} in setting $\delta_\mathrm{min}\! =\! 5$. For illustration,
the curvature profile $\langle\bar{d}_\mathrm{av}(\delta)/\delta\rangle_{N_2}$ for $N_2\! =\! 40$k is shown in Fig.\ \ref{fig:CPcomb40k}, together
with the curvature profile of the four-sphere that best fits the data, where only data points with $\delta_\mathrm{min}\! <\! \delta\! \leq\! 15 $ are included in the fit. 
It closely resembles Fig.\ 4 of reference \cite{qrc2}
for the same quantity, with the exception of the data point at $\delta\! =\! 1$.\footnote{We believe that the value at $\delta\! =\! 1$ in \cite{qrc2} may be incorrect, perhaps due
to a normalization error when implementing formula (\ref{sdistdis}), which affects measurements for small $\delta$, see \cite{Niestadt} for further discussion. \label{foot10}}
To take the nonuniversal character of $c_\mathrm{av}$ into account, a vertical shift has been applied to the continuum curve to make it
coincide with the data point at $\delta_\mathrm{min}$. The remaining fit parameter is the curvature radius $\rho$ of the continuum four-sphere. 
We will call $\rho_\mathrm{eff}$ the best fit for the curvature radius for a given sphere dimension $D$ and discrete volume $N_2$. 
For the case at hand, with $D\! =\! 4$ and $N_2\! =\! 40$k, we have found $\rho_\mathrm{eff}\! =\! \num{16.67 \pm 0.07}$, which is compatible 
with the value $\rho_\mathrm{eff}\! =\! \num{16.71 \pm 0.09}$ obtained in \cite{qrc2}. 

Because of the similarity of the curvature profiles for spheres of different dimension $D$ (cf.\ Fig.\ \ref{fig:dspheres}), it is not possible to determine with any 
certainty the optimal $D$ merely based on the quality of fits like that shown in Fig.\ \ref{fig:CPcomb40k}. 
This issue was addressed in \cite{qrc2} by invoking the following additional consistency argument. 
As a function of the curvature radius $\rho$, the volume $V^{({\cal D})}$ of a continuum $\cal D$-sphere scales like
\begin{equation}
V^{({\cal D})}\propto \rho^{\cal D}.
\label{volscale1}
\end{equation}

\begin{figure}[t]
\centering
\includegraphics[width=0.5\textwidth]{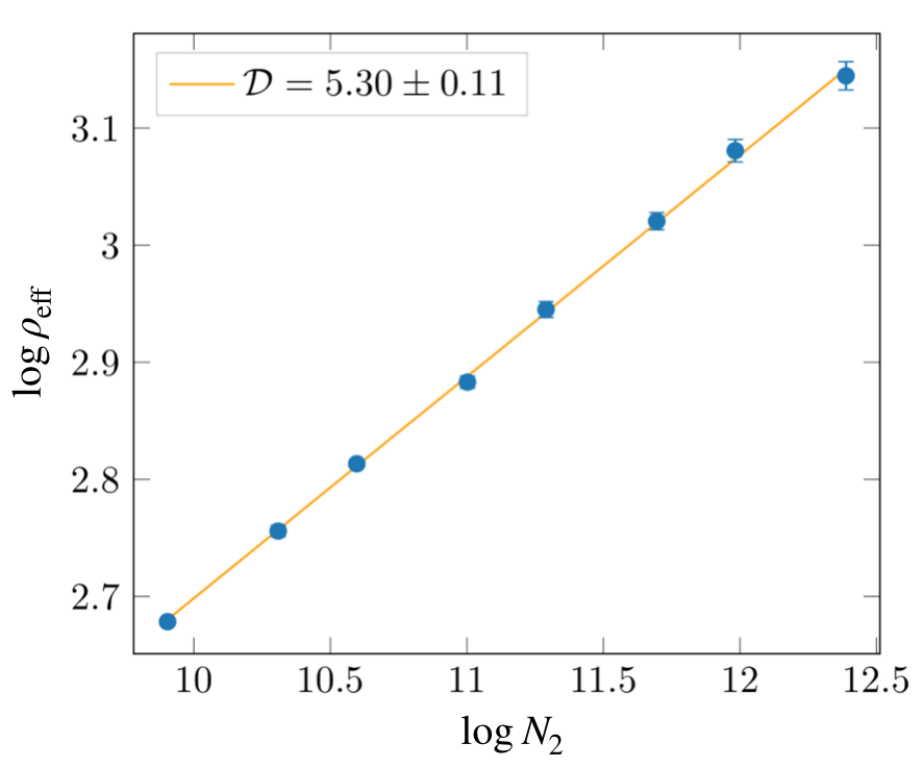}
\caption{Effective curvature radius $\rho_\mathrm{eff}$ as a function of the two-volume $N_2$ in a log-log plot for $D\! =\! 4$ and $\delta_\mathrm{min}\! =\! 5$, 
in the combinatorial ensemble
${\cal T}_c$, together with a linear fit (yellow line) corresponding to a sphere scaling dimension ${\cal D}\! =\! \num{5.30 \pm 0.11}$. 
}
\label{fig:Dfromrho4}
\end{figure}
\noindent Although the (average) triangulations are clearly not exact spheres globally, as can be seen from the asymmetry of the curves for the shell volumes in
Fig.\ \ref{fig:distr3E}, the scaling ansatz (\ref{volscale1}) does not depend on these details and should be rather robust. 
Following \cite{qrc2}, we assume that the quantum geometry resembles a continuum $\cal D$-sphere globally, in the sense that its volume scales with the 
${\cal D}$th power of its curvature radius. The consistency of this (mild) assumption is shown below. 
It implies that for a given choice of $D\! =\! 2$, 3, 4 or 5, the dis\-crete volume $N_2$ is expected to scale with the $\cal D$th power
of the effective curvature radius,
\begin{equation}
N_2\propto \rho_\mathrm{eff}^{\cal D}.
\label{volscale2}
\end{equation}

\noindent We can then extract for each $D$ a \textit{sphere scaling dimension} $\cal D$ from a best fit of the data set 
$\{ (N_2,\rho_\mathrm{eff} (N_2))\}$ for $N_2\!\in\! [20\mathrm{k},240\mathrm{k}]$
to the functional form (\ref{volscale2}). 
The dimension $D$ that describes the curvature properties of 2D Euclidean quantum gravity best is the
one where $D$ and $\cal D$ agree, within error bars. For illustration, Fig.\ \ref{fig:Dfromrho4} shows a log-log plot of the effective curvature radius $\rho_\mathrm{eff}$
as a function of the two-volume $N_2$, for $D\! =\! 4$. Within error bars, all points fall on a straight line, corresponding to a linear fit with ${\cal D}\! =\! \num{5.30 \pm 0.11}$.
We have repeated the same measurements and fitting procedure for $D\! =\! 2$, 3 and 5, resulting in fits of a similar quality.

\begin{table}[h]
\centering
    \begin{tabular}{
            l
            S[table-format=1.2,
            table-figures-uncertainty=1]
            S[table-format=1.2,
            table-figures-uncertainty=1]
        }
        \toprule
        \multicolumn{1}{c}{} & \multicolumn{2}{c}{$\mathcal{D}$} \\ \cmidrule(r){2-3}
        {$D$} & {present results} & {results from \cite{qrc2}} \\
        \midrule
        2  & 6.00\pm 0.14 & 5.70 \pm 0.30   \\
        3  & 5.38\pm 0.12 & 5.02 \pm 0.17 \\
        4  & 5.30\pm 0.11 & 4.92 \pm 0.17 \\
        5  & 5.15 \pm 0.11 & 4.85 \pm 0.16 \\
        \bottomrule
    \end{tabular}
\caption{Best sphere fitting dimension $\cal D$, extracted from the scaling ansatz (\ref{volscale2}), for integer dimension $D\!\in\! [2,5]$, in the combinatorial ensemble,
and for $\delta\!\in\! [5,15]$.}
\label{table:ddr}
\end{table}

The outcomes are collected in Table \ref{table:ddr}, next to the corresponding results found in reference \cite{qrc2}. The latter are systematically smaller than ours by at least one
standard deviation. Since details of the derivation of \cite{qrc2} were no longer available, we performed multiple cross-checks of our implementation in search of 
possible errors, but did not find any. The normalization error mentioned in
footnote \ref{foot10} would lead to smaller values of $\cal D$, like those reported in \cite{qrc2}, and may explain at least 
part of the discrepancy with our estimates \cite{Niestadt}. Despite
leaving this issue open, our conclusion from the results in Table \ref{table:ddr} for the combinatorial ensemble is identical to that reached in \cite{qrc2}, namely that
$\cal D$ and $D$ match best when $D\! =\! 5$. 

\begin{figure}[t]
\centering
\includegraphics[width=1.0\textwidth]{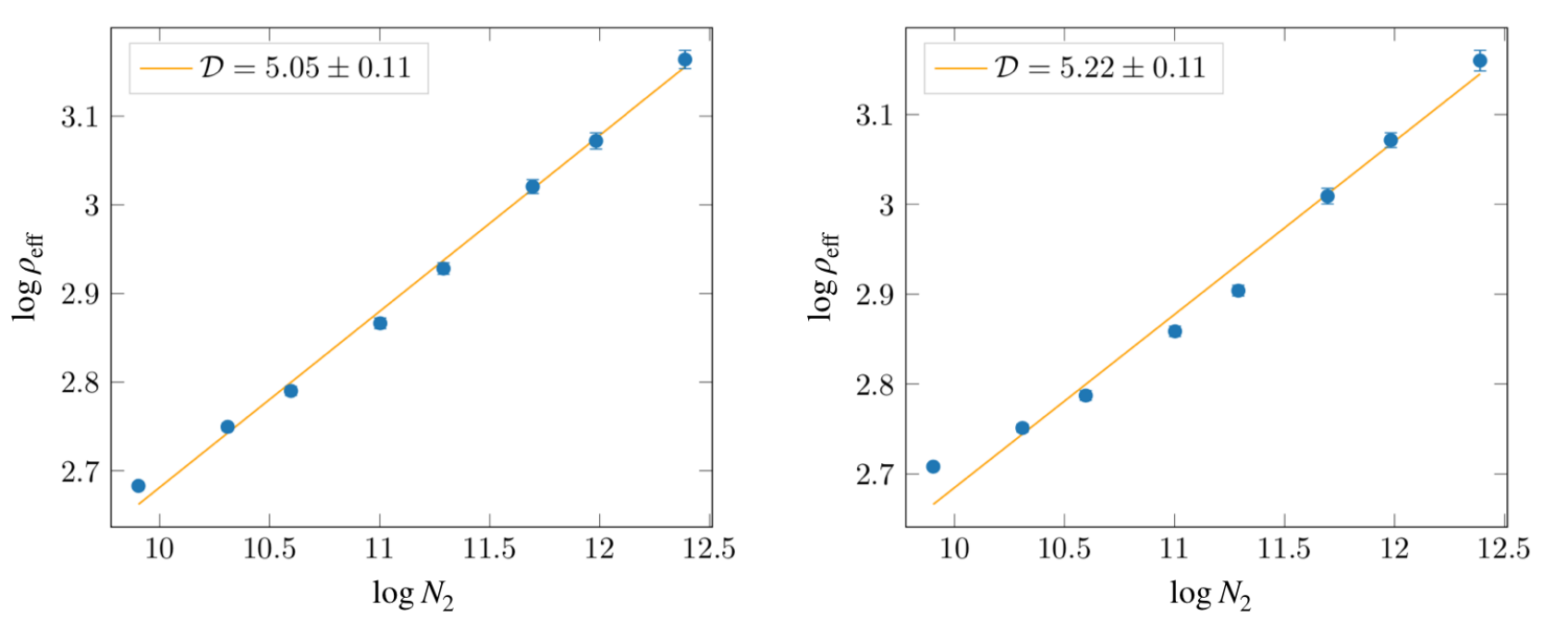}
\caption{Effective curvature radius $\rho_\mathrm{eff}$ as a function of the two-volume $N_2$ in a log-log plot for $D\! =\! 4$ and $\delta_\mathrm{min}\! =\! 5$, 
in the restricted degenerate ensemble
${\cal T}_r$ (left) and the maximally degenerate ensemble ${\cal T}_m$ (right), together 
with linear fits corresponding to the sphere scaling dimensions ${\cal D}\! =\! \num{5.05 \pm 0.11}$ and ${\cal D}\! =\! \num{5.22 \pm 0.11}$. 
}
\label{fig:Dfromrho_2}
\end{figure}
However, repeating the same set of measurements on the degenerate ensembles ${\cal T}_r$ and ${\cal T}_m$ throws new light on the (non-)reliability of the
data for small $N_2$ measured in the ${\cal T}_c$-ensemble. For illustration, we again show log-log plots of the effective curvature radius $\rho_\mathrm{eff}$ as a function of the volume $N_2$
for the case $D\! =\! 4$ (Fig.\ \ref{fig:Dfromrho_2}), while the best sphere fitting dimensions ${\cal D}$ for all $D$ are collected in Table \ref{table:ddrest}. 
\begin{table}[b]
\centering
    \begin{tabular}{
            l
            S[table-format=1.2,
            table-figures-uncertainty=1]
            S[table-format=1.2,
            table-figures-uncertainty=1]
        }
        \toprule
        \multicolumn{1}{c}{} & \multicolumn{2}{c}{$\mathcal{D}$} \\ \cmidrule(r){2-3}
        {$D$} & {${\cal T}_r$} & {${\cal T}_m$} \\
        \midrule
        2  & 5.76 \pm 0.13 & 5.97 \pm 0.14 \\
        3  & 5.14 \pm 0.11 & 5.31 \pm 0.12 \\
        4  & 5.05 \pm 0.11 & 5.22 \pm 0.11 \\
        5  & 4.91 \pm 0.10 & 5.07 \pm 0.12 \\
        \bottomrule
    \end{tabular}
\caption{Best sphere fitting dimension $\cal D$ for integer dimension $D\!\in\! [2,5]$, in the restricted
and maximally degenerate ensembles, and for $\delta\!\in\! [5,15]$.}
\label{table:ddrest}
\end{table}
Similar to the results for the combinatorial ensemble, $\cal D$ decreases with increasing $D$, and both dimensions match best at $D\! =\! 5$. 
Before taking this as a confirmation of the conclusion of \cite{qrc2}, we note that there is a significant difference with the combinatorial ensemble. In contrast to the apparently 
good fit of Fig.\ \ref{fig:Dfromrho4}, it is clear from Fig.\ \ref{fig:Dfromrho_2} that for the other two ensembles a power law is not a good fit for the data, although
our Hausdorff dimension measurements suggested that these ensembles should suffer less from lattice effects. We have observed this behaviour not only for
$D\! =\! 4$, but also for the other values of $D$. Roughly speaking, the slope is flatter for the three smallest triangulation sizes, which 
may signal the presence of finite-size effects. Alerted by these findings, we have performed a detailed analysis of discretization and finite-size effects, which
will be presented in the following subsections.

\subsubsection{Determining \boldmath$\delta_\mathrm{min}$}
\label{sub2}

Discretization artefacts are known to affect results for the curvature profile for small lattice distances $\delta$. One characteristic feature that seems to be
present generically in measurements of this quantity, 
including in the well-controlled case of regular flat lattices \cite{qrc1}, is an initial overshoot for small $\delta$. It is largest for $\delta\! =\! 1$ and 
then falls steeply toward a curve that is associated with genuine continuum behaviour. This overshoot region is clearly visible in the measured data shown in
Fig.\ \ref{fig:CPcomb40k}. The subtlety in cases where the continuum curvature profile is not known a priori, like the present analysis, is to determine the lower
bound $\delta_\mathrm{min}$ below which data points should be discarded. As explained at the beginning of Sec.\ \ref{sec:curv}, this determines the quantity
$c_\mathrm{av}$ and subsequent curve fits, and therefore potentially affects the value of the effective curvature radius $\rho_\mathrm{eff}$ extracted from it. 
A standard choice is $\delta_\mathrm{min}\! =\! 5$, which was used in \cite{qrc2} and in our re-analysis described in the previous subsection. 
\begin{figure}[t]
\centering
\includegraphics[width=0.52\textwidth]{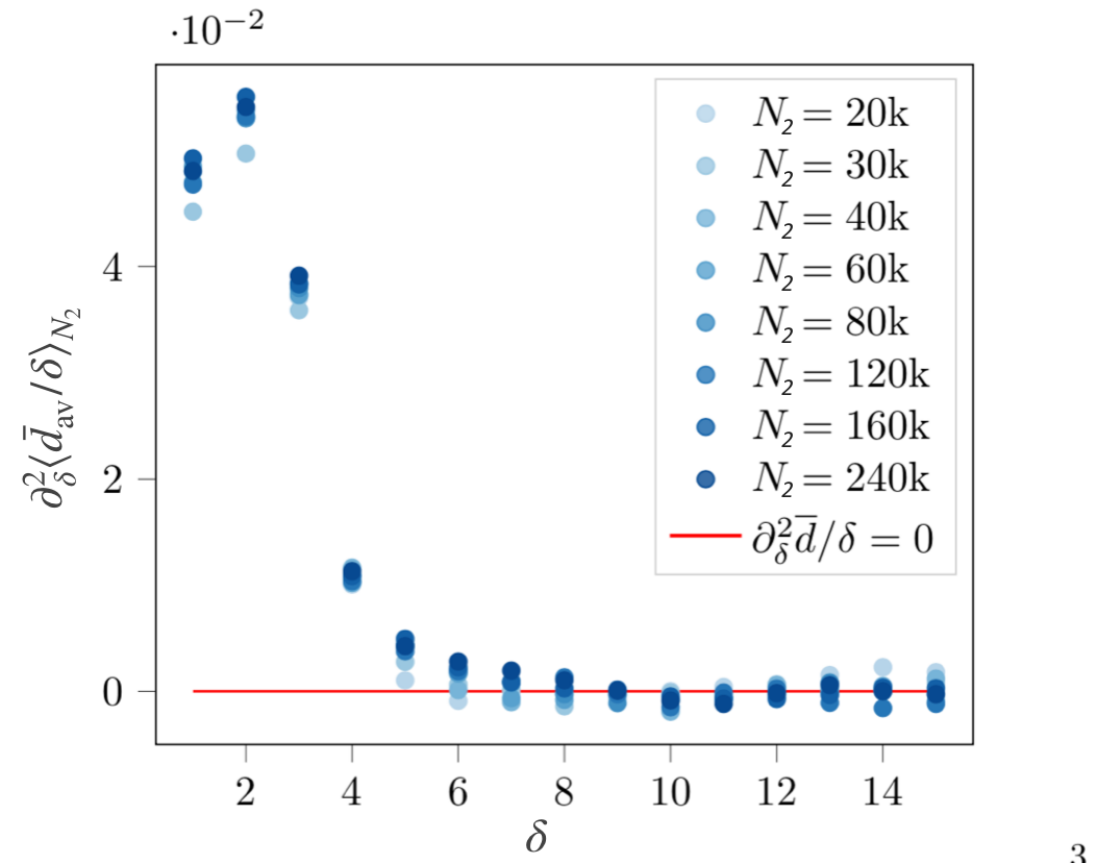}
\caption{Discrete second derivative of the curvature profile in the combinatorial ensemble, for two-volumes $N_2\! \in\! [20\mathrm{k},240\mathrm{k}]$. 
The horizontal line indicates
$\partial^2_\delta \langle\bar{d}_\mathrm{av}/\delta\rangle_{N_2}\! = \! 0$. For clarity of presentation, error bars (ranging between 0.001 and 0.005)
have been omitted.}
\label{fig:deriv1}
\end{figure}

Instead of relying purely on a visual inspection of the data points, we will determine $\delta_\mathrm{min}$ by a quantitative criterion, namely
as the transition point between a ``convex range" for small $\delta$
(associated with discretization artefacts and best described by some convex function of $\delta$) and a ``concave range" (associated with the continuum curvature profile
of some $D$-dimensional sphere and described by a concave function of $\delta$). To this end, we compute the (discrete) second derivative 
$\partial^2_\delta \langle\bar{d}_\mathrm{av}/\delta\rangle_{N_2}$ of the 
curvature profile $\langle\bar{d}_\mathrm{av}/\delta\rangle_{N_2}$ with respect to $\delta$ for each given volume $N_2$.\footnote{The discrete 
second derivative is obtained by two consecutive applications of the \textit{numpy.gradient} function of the NumPy library \cite{numpy}.}
We will take its vanishing as an indicator of the searched-for transition point, and choose the lower bound $\delta_\mathrm{min}$ to be at or near this point.
Since discrete derivatives are susceptible to noise, this criterion is still not rigorous, but in our view represents an improvement on the previous method.
This view will be corroborated further by the outcome of our analysis below.   

\begin{figure}[t]
\centering
\includegraphics[width=1.0\textwidth]{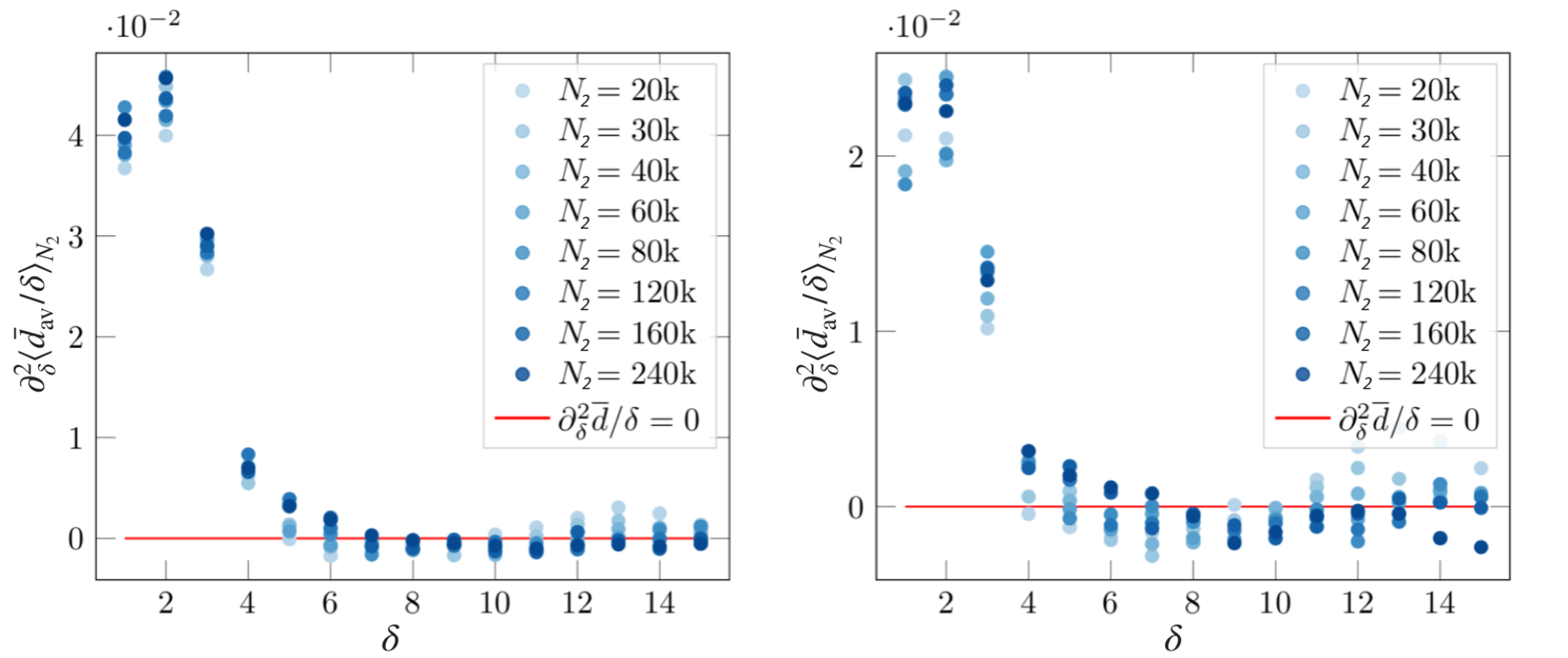}
\caption{Discrete second derivative of the curvature profile in the restricted degenerate ensemble (left) and the maximally degenerate ensemble (right), 
for two-volumes $N_2\! \in\! [20\mathrm{k},240\mathrm{k}]$. The horizontal lines indicate
$\partial^2_\delta \langle\bar{d}_\mathrm{av}/\delta\rangle_{N_2}\! = \! 0$. Error bars, ranging between 0.001 and 0.005,
have been omitted.}
\label{fig:deriv2}
\end{figure}

For the combinatorial ensemble ${\cal T}_c$ and triangulation sizes in the range $N_2\! \in\! [20\mathrm{k},240\mathrm{k}]$, 
the second derivative is shown in Fig.\ \ref{fig:deriv1}. 
The fact that for $\delta\! =\! 5$ most points are still in positive territory suggests that we should choose $\delta_\mathrm{min}\! >\! 5$. However, 
a larger $\delta_\mathrm{min}$ implies more discarded data points. Taking this trade-off into account, we have settled on $\delta_\mathrm{min}\! =\! 7$ 
as the best choice for the combinatorial ensemble. The second derivatives of the curvature profiles for the degenerate ensembles are depicted in
Fig.\ \ref{fig:deriv2}, from which we have extracted $\delta_\mathrm{min}\! =\! 6$ in either case as a sufficiently large lower bound. These values for 
$\delta_\mathrm{min}$ will be used from now on, including in Sec.\ \ref{sub3} below.

\begin{figure}[t]
\centering
\includegraphics[width=0.55\textwidth]{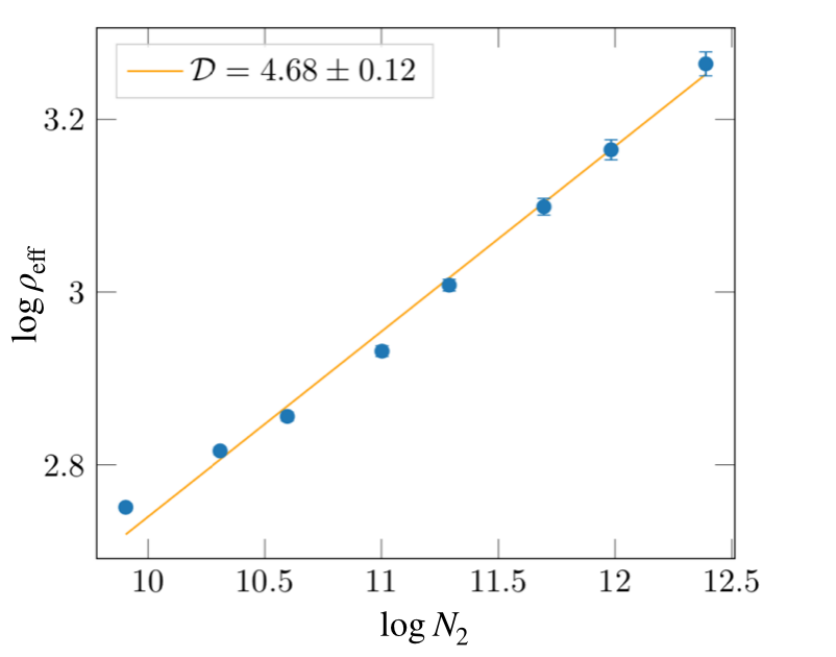}
\caption{Effective curvature radius $\rho_\mathrm{eff}$ as a function of the two-volume $N_2$ in a log-log plot for $D\! =\! 4$ and $\delta_\mathrm{min}\! =\! 7$, 
in the combinatorial ensemble
${\cal T}_c$. All data points are taken into account for determining the best linear fit (yellow line), corresponding to a sphere scaling dimension 
${\cal D}\! =\! \num{4.68 \pm 0.12}$.  
}
\label{fig:logimprove}
\end{figure}

Using the newly determined values for $\delta_\mathrm{min}$ alters the effective curvature radii $\rho_\mathrm{eff}$ obtained from the fits to the 
sphere curvature profiles for all three ensembles. Of particular interest are the changes for the combinatorial ensemble, where we found a significant quantitative difference 
compared to the results for $\delta_\mathrm{min}\! =\! 5$. Where in the previous log-log plot (Fig.\ \ref{fig:Dfromrho4}) of $\rho_\mathrm{eff}$ as a function of the volume $N_2$ the
data points fitted almost perfectly to a straight line, this is no longer true for the choice $\delta_\mathrm{min}\! =\! 7$. As can be seen in Fig.\ \ref{fig:logimprove}, 
the picture is now similar to what we found previously for the degenerate ensembles (Fig.\ \ref{fig:Dfromrho_2}), with a flatter slope for the points at small volume.
As a consequence, the values for $\cal D$ obtained from a power law fit are significantly smaller. For example, for $D\! =\! 4$, we obtained 
${\cal D}\! =\! \num{4.68 \pm 0.12}$, in contrast to our earlier finding ${\cal D}\! =\! \num{5.30 \pm 0.11}$. 
The observed deviations from a straight line are due to finite-size effects, as we will discuss in the following subsection.

\subsubsection{Finite-size effects}
\label{sub3}

The potential presence of significant finite-size effects for the smaller volumes is also supported by the following argument. 
Individual point pair distances $\bar{d}(q,q')$ contributing to the average sphere distance $\bar{d}(S_p^\delta,S_{p'}^\delta)$ of eq.\ (\ref{sdistdis}) 
have an upper bound of $3\delta$. With $\delta_\mathrm{max}\! =\! 15$, which we have used until now, the largest possible distance $\bar{d}(q,q')$ is given by 45. 
However, whenever $\delta$ is too large for a given volume $N_2$, the curvature profile at value $\delta$ probes global and topological features
of the underlying geometries, rather than the (quasi-)local curvature properties we are after. In addition, we already know that on sufficiently large linear scales 
the quantum geometry is not approximated well by a round sphere. As we noted in Sec.\ \ref{sec:haus}, the
plots of the average sphere volume as a function of the radius $r$ are not symmetric around their maxima as would be the case
for perfect spheres. 

In other words, our curvature construction and comparison with the curvature profile
of continuum spheres is only valid as long as the ``effective" linear size of the double sphere configuration of Fig.\ \ref{fig:qrcpic} is not too large, 
compared to the linear size of the entire geometry. 
\begin{figure}[t]
\centering
\includegraphics[width=1.01\textwidth]{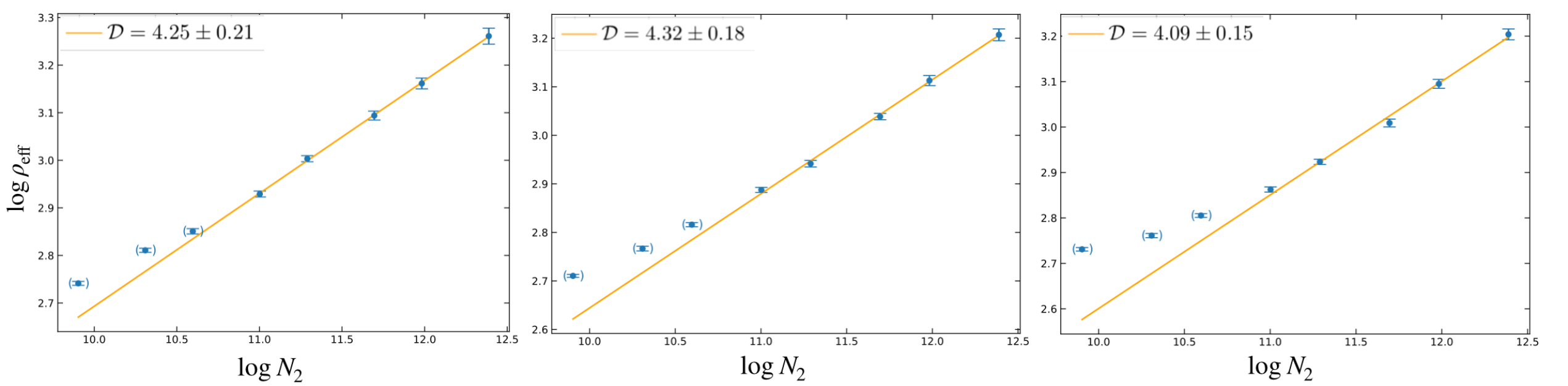}
\caption{Effective curvature radius $\rho_\mathrm{eff}$ as a function of the two-volume $N_2$ in a log-log plot for $D\! =\! 4$, 
in the reduced volume range $N_2\!\in\! [60\mathrm{k},240\mathrm{k}]$, for the combinatorial, restricted degenerate and maximally 
degenerate ensembles (left, middle and right respectively). Data points for lower-lying
volumes are shown in parentheses, but have not been included in the best linear fits (yellow lines). 
}
\label{fig:newloglog}
\end{figure}
To address this issue, we start by implementing a relatively crude method where we eliminate the volumes deemed too small to evade
significant finite-size effects, while keeping the choice $\delta_\mathrm{max}\! =\! 15$ fixed. 

\begin{table}[b]
    \centering

    \begin{tabular}{
            l
            S[table-format=1.2,
            table-figures-uncertainty=2]
            S[table-format=1.2,
            table-figures-uncertainty=2]
            S[table-format=1.2,
            table-figures-uncertainty=2]
        }
        \toprule
        \multicolumn{1}{c}{} & \multicolumn{3}{c}{$\mathcal{D}$} \\ \cmidrule(r){2-4}
        {$D$} & {${\cal T}_c$} & {${\cal T}_r$} & {${\cal T}_m$} \\
        \midrule
        2  & 4.54 \pm 0.22 & 4.66 \pm 0.20 & 4.44 \pm 0.17 \\
        3  & 4.29 \pm 0.21 & 4.36 \pm 0.17 & 4.12 \pm 0.15 \\
        4  & 4.25 \pm 0.21 & 4.32 \pm 0.18 & 4.09 \pm 0.15 \\
        5  & 4.21 \pm 0.22 & 4.24 \pm 0.19 & 4.02 \pm 0.16 \\
        \bottomrule
    \end{tabular}
      
\caption{Best sphere fitting dimension $\cal D$ for integer dimension $D\!\in\! [2,5]$ and all three ensembles.
Contrary to the values given in the earlier Tables \ref{table:ddr} and \ref{table:ddrest}, the volume range for the power-law fits has been restricted to $N_2\!\geq\! 60$k. 
For the ensembles ${\cal T}_c$, ${\cal T}_r$ and ${\cal T}_m$, $\delta_\mathrm{min}$ was set to 7, 6 and 6 respectively.
}
\label{table:hard}
\end{table}
In the absence of any unique or sharp criteria, we work with the requirements that
(i) configurations with a maximal diameter 45 should ``fit inside" the sphere-like geometry in the sense that $3\delta_\mathrm{max}\!\equiv\! 45 \! \lesssim\! r_\mathrm{max}$,
where the average maximal linear extension $r_\mathrm{max}$ for a given ensemble and volume $N_2$ is read off from Fig.\ \ref{fig:distr3E}, and (ii) the data points
of the included volumes should lie approximately on a straight line in the log-log plots like those shown in Figs.\ \ref{fig:Dfromrho_2} and \ref{fig:logimprove}. 
This leads to the exclusion of the effective curvature radii $\rho_\mathrm{eff}$ for the system sizes $N_2\! <\! 60$k for all three ensembles, corresponding to the three lowest-lying 
data points with $\log N_2\! <\! 11.0$ in the log-log plots. 
The resulting power-law fits for the particular case $D\! =\! 4$ are shown in Fig.\ \ref{fig:newloglog}. 
Unsurprisingly, the reduced volume range leads to a much better quality of the
linear fits, which also exhibit clearly that the omitted data points are associated with a flatter slope. The best sphere fitting dimensions $\cal D$ 
extracted from the scaling ansatz (\ref{volscale2}) for all ensembles are collected in Table \ref{table:hard}. A significant change has occurred in comparison with
the fits that included all volumes: for all three ensembles, the dimensions $\cal D$ and $D$ now agree best for $D\! =\! 4$, at the expense of somewhat larger error bars. 

\begin{figure}[t]
\centering
\includegraphics[width=1.01\textwidth]{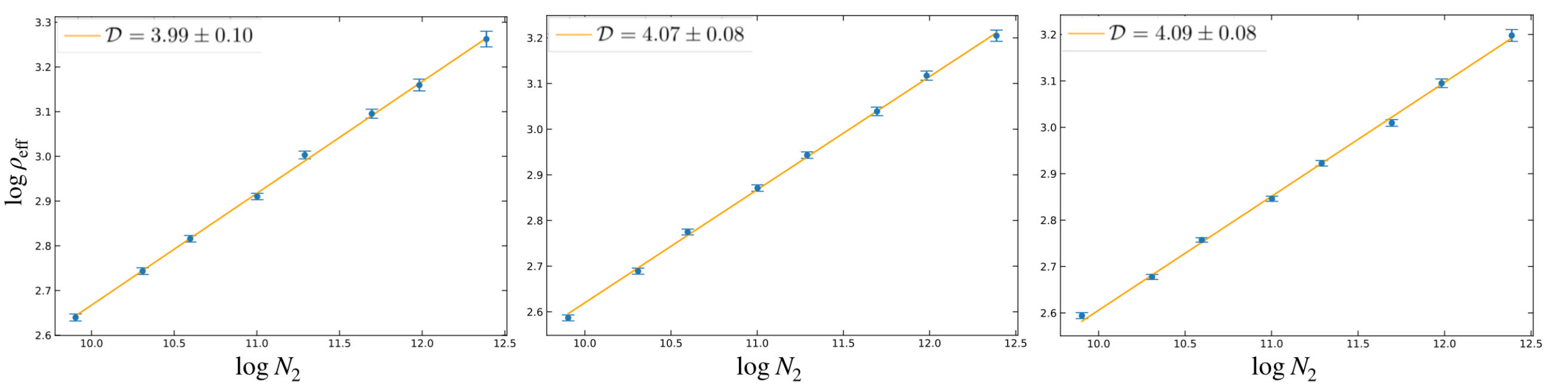}
\caption{Effective curvature radius $\rho_\mathrm{eff}$ as a function of the two-volume $N_2$ in a log-log plot for $D\! =\! 4$ and the full volume range 
$N_2\!\in\! [20\mathrm{k},240\mathrm{k}]$, 
but with data points for $\delta\! >\! \delta_\mathrm{max}(N_2)$ omitted, as described in the text.
Plots for the combinatorial, restricted degenerate and maximally degenerate ensembles (left, middle and right respectively). 
Best linear fits are indicated by the yellow lines.
}
\label{fig:logfinal}
\end{figure}

By way of further improvement, we implement next a more refined and less wasteful method to extract the sphere fitting dimensions. Instead of excluding all data points
for the volumes $N_2\! =\! 20$, 30 and 40k, we omit only those points at small volumes whose $\delta$ lies above 
a specific, \textit{$N_2$-dependent} threshold $\delta_\mathrm{max}(N_2)$.
To determine this maximal value, we select a reference volume $N_{2,\mathrm{ref}}\! =\! 80$k for all ensembles, 
for which -- by visual inspection of figures like Fig.\ \ref{fig:newloglog} -- the data point at $\delta_\mathrm{max}\! =\! 15$ is sufficiently unaffected by 
finite-size effects. We then make use of the scaling behaviour of the linear size as a function of the volume $N_2$ to fix the volume-dependent thresholds 
$\delta_\mathrm{max}(N_2)$ for $N_2\!\leq\! N_{2,\mathrm{ref}}$ according to
\begin{equation}
\delta_\mathrm{max}(N_2):=  \lceil 15 \cdot \big( N_2/N_{2,\mathrm{ref}}\big)^{1/d_H} \rceil,
\label{nscale}
\end{equation} 
where for simplicity we use the known value $d_H\! =\! 4$ for the Hausdorff dimension.\footnote{The same result is obtained when using the $d_H$-values 
in the last column of Table \ref{haustable}.}
The notation $\lceil\cdot\rceil$ in eq.\ (\ref{nscale}) denotes the ceiling function, which rounds off to the next higher integer. This leads to 
$\delta_\mathrm{max}(N_2)\! =\! 11$, 12, 13, and 14 for the smallest volumes $N_2\! =\! 20$, 30, 40 and 60k respectively, for all ensembles.
For volumes with $N_2\! \geq \! N_{2,\mathrm{ref}}$, we keep using $\delta_\mathrm{max}\! =\! 15$.

The resulting fits for $\log (\rho_\mathrm{eff})$ as a function of $\log (N_2)$, for $D\! =\! 4$ and all three ensembles, are presented in Fig.\ \ref{fig:logfinal}.
It is gratifying to see that our refined methodology appears to have eliminated successfully the finite-size effects for the smaller volumes, in the sense that
all data points now lie beautifully along straight lines. The results for the best sphere fitting dimension $\cal D$ extracted from the fits for all ensembles
are presented in Table \ref{table:final}. 

\begin{table}[t]
    \centering 
    \begin{tabular}{
            l
            S[table-format=1.2,
            table-figures-uncertainty=2]
            S[table-format=1.2,
            table-figures-uncertainty=2]
            S[table-format=1.2,
            table-figures-uncertainty=2]
        }
        \toprule
        \multicolumn{1}{c}{} & \multicolumn{3}{c}{$\mathcal{D}$} \\ \cmidrule(r){2-4}
        {$D$} & {${\cal T}_c$} & {${\cal T}_r$} & {${\cal T}_m$} \\
        \midrule
        2  & 4.19 \pm 0.09 & 4.27 \pm 0.08 & 4.31 \pm 0.08 \\
        3  & 4.03 \pm 0.09 & 4.09 \pm 0.08 & 4.12 \pm 0.08 \\
        4  & 3.99 \pm 0.10 & 4.07 \pm 0.08 & 4.09 \pm 0.08 \\
        5  & 3.99 \pm 0.09 & 4.04 \pm 0.09 & 4.05 \pm 0.08 \\
        \bottomrule
    \end{tabular}
 \caption{Best sphere fitting dimension $\cal D$ for integer dimension $D\!\in\! [2,5]$ and all three ensembles.
To eliminate finite-size effects, the $\delta$-ranges for the volumes $N_2\! =\! 20$, 30, 40 and 60k were limited by $\delta_\mathrm{max}(N_2)\! =\! 11$, 12, 13 and 14 respectively.
For the ensembles ${\cal T}_c$, ${\cal T}_r$ and ${\cal T}_m$, $\delta_\mathrm{min}$ was set to 7, 6 and 6 respectively.}
\label{table:final}
\end{table}

\begin{figure}[t]
\centering
\includegraphics[width=1.0\textwidth]{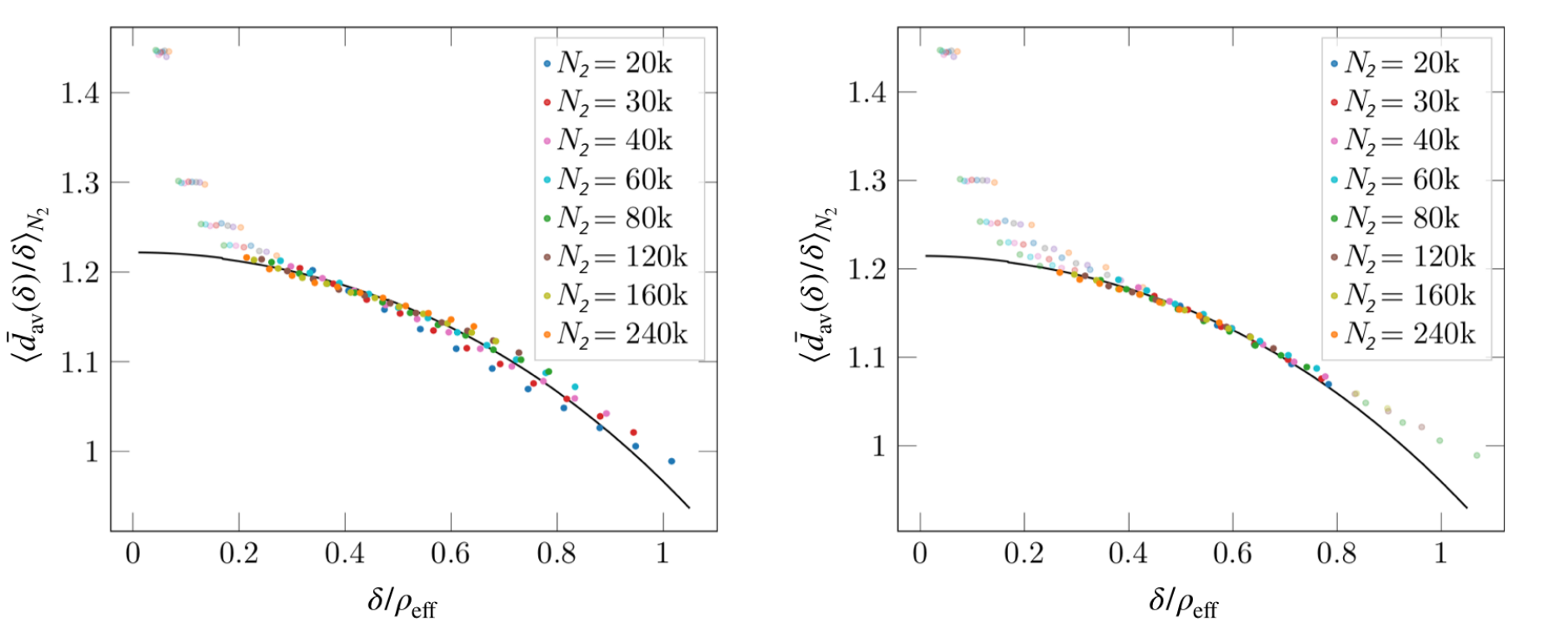}
\caption{Curvature profiles $\langle\bar{d}_\mathrm{av}(\delta)/\delta\rangle_{N_2}$ for the combinatorial ensemble, sphere di\-men\-sion $D\! =\! 4$
and volumes $N_2\! \in\! [20\mathrm{k},240\mathrm{k}]$, as a function of the rescaled dis\-tance $\delta /\rho_\mathrm{eff}$, 
exhibiting finite-size scaling. Results for $\delta_\mathrm{min}\! =\! 5$ and fixed $\delta_\mathrm{max}\! =\! 15$ (left), and
of our improved analysis with $\delta_\mathrm{min}\! =\! 7$ and variable $\delta_\mathrm{max}(N_2)$ according to eq.\ (\ref{nscale})
(right). The smooth curves are the curvature profiles of continuum four-spheres with a best-fit constant vertical shift.
Data points with $\delta\! <\! \delta_\mathrm{min}$ or $\delta\! >\! \delta_\mathrm{max}$ are shown in light colours, but have not been included in the fits.
}
\label{fig:fss}
\end{figure}

\noindent Compared with our previous method, which omitted all data points for volumes $N_2\! <\! 60$k, the error bars have been approximately halved. 
As a further improvement, for $D\! =\! 4$, the values for $D$ and $\cal D$ now coincide within error bars (for ${\cal T}_c$ and ${\cal T}_r$) or lie just outside
the error bar (for ${\cal T}_m$). The excellent quality of this agreement leads us to conclude that the local curvature properties of 2D Euclidean quantum gravity 
resemble closest those of a continuum sphere of dimension four, contrary to the earlier finding of five as the preferred dimension \cite{qrc2}.

To further illustrate the progress we have made relative to previous results, let us examine the ``curve collapse" of Fig.\ 6 in \cite{qrc2} for the combinatorial ensemble, 
which collects the curvature profiles $\langle\bar{d}_\mathrm{av}(\delta)/\delta\rangle_{N_2}$ as a function of the rescaled distances $\delta/\rho_\mathrm{eff}$ 
for all volumes $N_2$. If they fall on a common curve, it indicates the presence of finite-size scaling, with a well-defined infinite-volume limit. 
Fig.\ \ref{fig:fss} shows the corresponding graphs before and after the improvements we have described in this Section, together with continuum
curves for the curvature profile of a round four-sphere with a best vertical shift. 
We observe that in the plot on the right the overlap among data for different volumes and their closeness to the continuum curve is much improved,
for the subset of points considered, lending further support to our conclusion.

\section{Summary and outlook}
\label{sec:sum}

We set out to examine the curvature of 2D Euclidean quantum gravity, given as the continuum limit of a path integral over two-dimensional
dynamical triangulations. The curvature observable we considered was the curvature profile, defined as the expectation value of the
average distance of two $\delta$-spheres divided by $\delta$, eq.\ (\ref{avdis}). A previous investigation found that for sufficiently small $\delta$
this quantity can be fitted to the curvature profile of a classical round, $D$-dimensional sphere, with the best match to a sphere of dimension
$D\! =\! 5$ \cite{qrc2}. 

The curvature profiles for continuum spheres with $D\! =$ 2, 3, 4 and 5 are shown in Fig.\ \ref{fig:dspheres}. For these Riemannian manifolds of
constant, positive curvature the scale of the curvature is set by the curvature radius $\rho$ of the spheres, with all sectional curvatures being equal
to $1/\rho^2$. In the region $\delta/\rho\!\lesssim\! 1$, which we use for comparison with the measurements in the quantum gravity theory, the classical
curvature profiles are well described by their quadratic approximations,
\begin{equation}
\frac{\bar{d} (S^\delta_p,S^\delta_{p'})}{\delta} \approx c_q(D) - \tilde{c}(D) \bigg(\frac{\delta}{\rho}\bigg)^2  ,
\label{approx}
\end{equation}
where $c_q(D)>0$ and $\tilde{c}(D)\! >\! 0$ are dimension-dependent constants. As can be seen from Fig.\ \ref{fig:dspheres}, both constants increase slightly as 
functions of $D$ for the dimensions we
are interested in. This weak dependence makes it difficult to discriminate between different values
of $D$ when fitting the numerical data. Following \cite{qrc2}, we therefore employed an additional consistency argument by determining the scaling
behaviour of the effective curvature radius $\rho_\mathrm{eff}$ with the volume $N_2$ for a given initial choice of the dimension $D$ of the continuum
sphere whose curvature profile we are fitting to. Namely, we demanded that the parameter $\cal D$ that describes this scaling according to
$\rho_\mathrm{eff} \propto  N_2^{1/{\cal D}}$ should match the value of $D$. 

We began by measuring average sphere distances according to the specifications of \cite{qrc2}, i.e.\ in the ensemble ${\cal T}_c$ 
of combinatorial triangulations and using data points in the interval $\delta\!\in\! [5,15]$. At first glance this appeared to confirm the previous conclusion
that the preferred sphere dimension is $D\! =\! 5$, until we performed the same analysis for the two degenerate ensembles
${\cal T}_r$ and ${\cal T}_m$, which strongly suggested the presence of lattice artefacts. Re-examining this issue 
in all three ensembles and carefully selecting data points deemed unaffected by such artefacts resulted in fits of very good quality 
(cf.\ Fig.\ \ref{fig:logfinal}), an improved finite-size scaling (illustrated by Fig.\ \ref{fig:fss}) and a new and different conclusion: it is not for $D\! =\! 5$ but 
for $D\! =\! 4$ that the two dimensions $D$ and $\cal D$ coincide, and they do so within error bars.

The fact that the (expectation value of the) curvature radius $\rho_\mathrm{eff}$ scales like $N_2^{1/4}$, which coincides with the (noncanonical) scaling behaviour
of distances $\delta\!\propto\! N_2^{1/d_H}\!\equiv\! N_2^{1/4}$ in 2D Euclidean quantum gravity implies that in the conti\-nuum limit\footnote{Equivalently,
the continuum limit can be defined as $a\rightarrow 0$, where by assumption the lattice spacing $a$ is related to the lattice volume $N_2$ via 
$V_2\! =\! a^2 N_2$, for a fixed, dimensionful two-volume $V_2$.} $N_2\rightarrow\infty$ 
their ratio $\delta /\rho_\mathrm{eff}$ can converge to a finite, nonvanishing number. 
Recall that according to relation (\ref{avcurv}) the nonconstant part of the curvature profile -- up to a negative constant -- is the 
dimensionless average quantum Ricci scalar $K_\mathrm{av}$. Our results suggest that for sufficiently small lattice distances $\delta$, 
this quantity has a well-defined continuum limit
\begin{equation}
K_\mathrm{av}\sim \bigg( \frac{\delta}{\rho_\mathrm{eff}} \bigg)^2     \xrightarrow{N_2\rightarrow\infty}
\bigg( \frac{\delta_\mathrm{ph}}{\rho_\mathrm{ph}} \bigg)^2 \sim \delta_\mathrm{ph}^2 K_\mathrm{ph},
\label{contcurv}
\end{equation}
where to leading order in $a$, the physical, dimensionful variables $\delta_\mathrm{ph}$ and $\rho_\mathrm{ph}$ depend on their lattice
counterparts according to $\delta_\mathrm{ph}\! :=\! \delta \sqrt{a}$ and $\rho_\mathrm{ph}\! :=\! \rho_\mathrm{eff}\sqrt{a}$, and
$K_\mathrm{ph}$ is the physical average quantum Ricci scalar of the continuum theory. The noncanonical scaling of $\delta_\mathrm{ph}$ implies
that also $K_\mathrm{ph}$ scales noncanonically. What this implies for its role and interpretation in the quantum theory remains a matter for
further research.

It raises the interesting question of whether and how this or other notions of quantum curvature can be derived analytically. So far, despite a resurgence of interest
in 2D Euclidean quantum gravity by physicists, mathematical physicists and pure mathematicians \cite{Mertens2020,Budd2022,Sheffield}, the
definition and construction of a curvature inherent to the quantum theory has received little attention. 
An exception is the recent study of a notion of Gaussian curvature in Liouville quantum gravity as the scaling limit
of a deficit-angle prescription, using a particular random planar map model \cite{hip}. 
As we have already pointed out in the introduction, if a notion of quantum curvature exists, as our work suggests, it is far
from clear that it should be unique. 
We hope that our nonperturbative numerical analysis of the quantum Ricci curvature can encourage researchers in this and other approaches 
to 2D Euclidean quantum gravity to have a closer look at curvature, where much remains to be discovered and understood.

\vspace{0.2cm}
\subsubsection*{Acknowledgments} 

RL thanks the Perimeter Institute for hospitality.
This research was supported in part by Perimeter Institute for Theoretical Physics. Research at Perimeter Institute is supported by the 
Government of Canada through the Department of Innovation, Science and Economic Development and by the Province of Ontario through
the Ministry of Colleges and Universities. 

\vspace{0.4cm}
\subsection*{Appendix}

In this appendix, we define the three ensembles in terms of their geometric properties as (generalized) triangulations, which is how they appear
in the numerical investigations. We describe in more detail the geometric irregularities of triangulations associated with tadpoles and self-energy subgraphs in the dual
graph picture, and how these different characterizations relate to each other. Although bits and pieces are covered and referred to in the literature (see e.g.\ 
\cite{ACMbook,thor,Ambjorn1997}), we have not found a place where all of these aspects are discussed together. 

Our starting point are finite triangulations with the topology of a two-sphere $S^{(2)}$, 
whose elementary building blocks are flat equilateral 
triangles (two-simplices $\sigma^{(2)}$) of edge length $a$, 
the already mentioned UV-cutoff. $D$-simplices $\sigma^{(D)}$ of lower dimension $D$ contained in the triangles are their edges or links (one-simplices $\sigma^{(1)}$) 
and vertices (zero-simplices $\sigma^{(0)}$). 
The ensembles differ by how the triangles can be glued together pairwise along their edges to obtain a piecewise flat geometry $T$. 
\begin{figure}[t]
\centering
\includegraphics[width=0.4\textwidth]{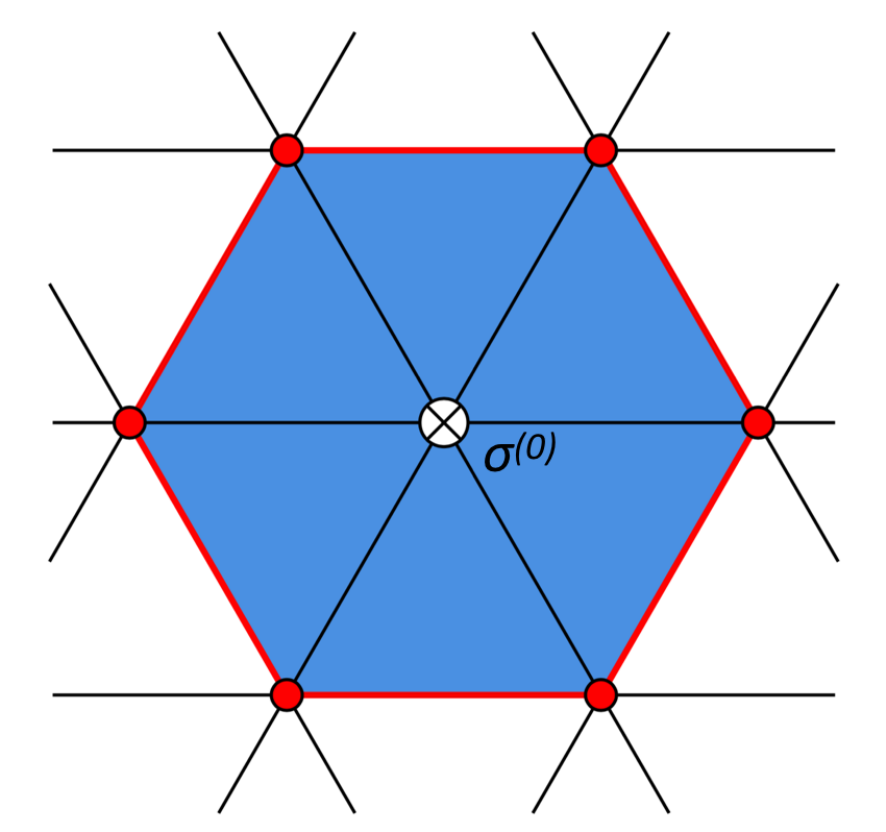}
\caption{The star (blue) and link (red) of a zero-simplex $\sigma^{(0)}$, the vertex at the centre (marked with a cross).
The link shown here is homeomorphic to $S^{(1)}$.
}
\label{fig:star}
\end{figure}
In what follows we will need a few definitions regarding simplicial structures \cite{ACMbook,Gallier}. 
A \textit{face} of a simplex $\sigma$ is any simplex whose vertices are a subset of the
vertices of $\sigma$. The \textit{star} st$(\sigma)$ of a simplex $\sigma$ in a triangulation $T$ is the union of all simplices in $T$ that have $\sigma$ as a face. Lastly, the
\textit{link} lk$(\sigma)$ of a simplex $\sigma$ is the union of all simplices $\tau\!\in\! \mathrm{st}(\sigma)$ which have vanishing intersection with $\sigma$, 
$\tau\cap\sigma\! =\! \emptyset$. 
\begin{figure}[H]
\centering
\includegraphics[width=0.95\textwidth]{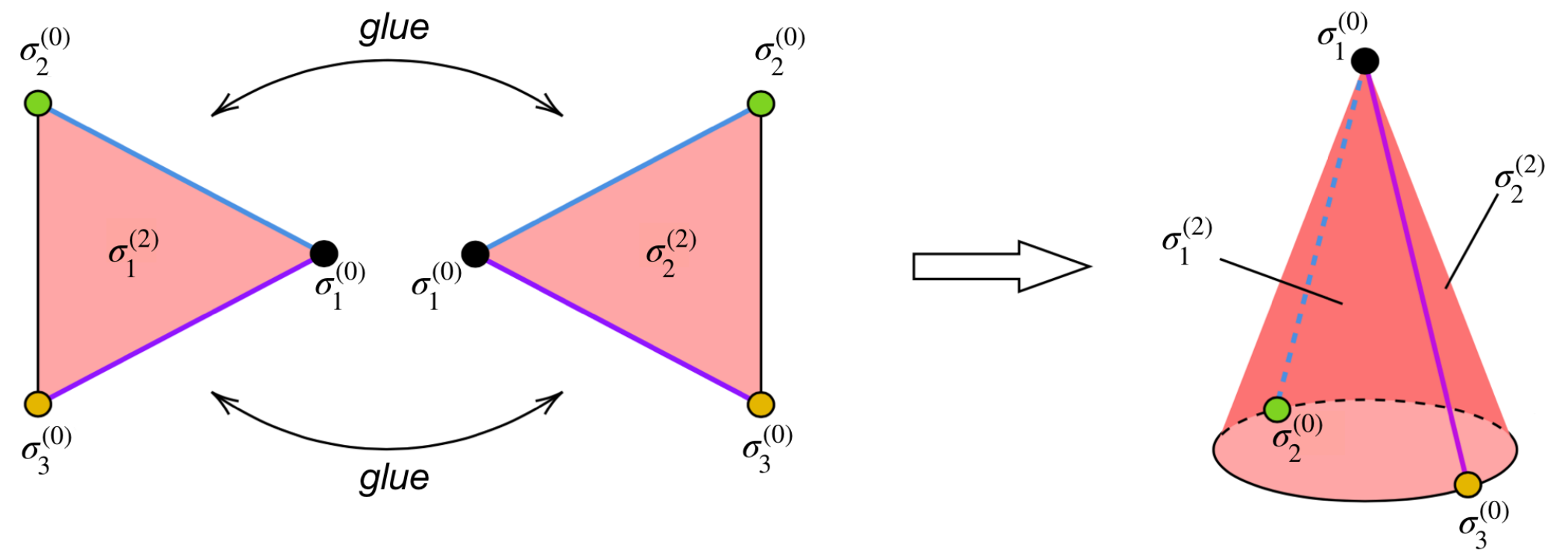}
\caption{The vertices $\sigma^{(0)}_1$, $\sigma^{(0)}_2$ and $\sigma^{(0)}_3$ are all distinct, but if we glue the triangles 
$\sigma^{(2)}_1$ and $\sigma^{(2)}_2$ on the left together as indicated,
the vertices belong to both of them. The resulting conical ``hat" on the right cannot be part of a combinatorial triangulation, 
but is allowed as part of a restricted degenerate triangulation. 
}
\label{fig:hat}
\end{figure}
\noindent Fig.\ \ref{fig:star} shows an example of the star and link of a zero-simplex $\sigma^{(0)}$.\footnote{The notion of ``link" introduced here
should not be confused with what elsewhere we use as a synonym for ``edge".}
A triangulation $T$ of $S^{(2)}$ is combinatorial, $T\!\in\! {\cal T}_c$, if each of its triangles and edges is combinatorially distinct. 
By this we mean that each triangle is
defined uniquely by a set of three distinct vertices, and each edge by a set of two distinct vertices. 
This excludes e.g.\ the presence of two distinct edges that
share the same vertices, or of two distinct triangles that have more than one edge in common. 
The graph dual to a 2D combinatorial triangulation
is a trivalent $\phi^3$-graph without tadpole or self-energy subgraphs \cite{thor}.   
Yet another way to characterize elements of ${\cal T}_c$ uses the definitions just introduced: a triangulation is
combinatorial if the link of each zero-simplex is homeomorphic to a one-sphere (cf.\ Fig.\ \ref{fig:star}), and the link of each one-simplex is homeomorphic to a zero-sphere (the set of two points),
\begin{equation}
\mathrm{lk}(\sigma^{(0)}) \cong S^{(1)}\;\;\;\;\;\mathrm{and}\;\;\;\;\; \mathrm{lk}(\sigma^{(1)}) \cong S^{(0)}.
\label{conds}
\end{equation}
\begin{figure}[t]
\centering
\includegraphics[width=0.48\textwidth]{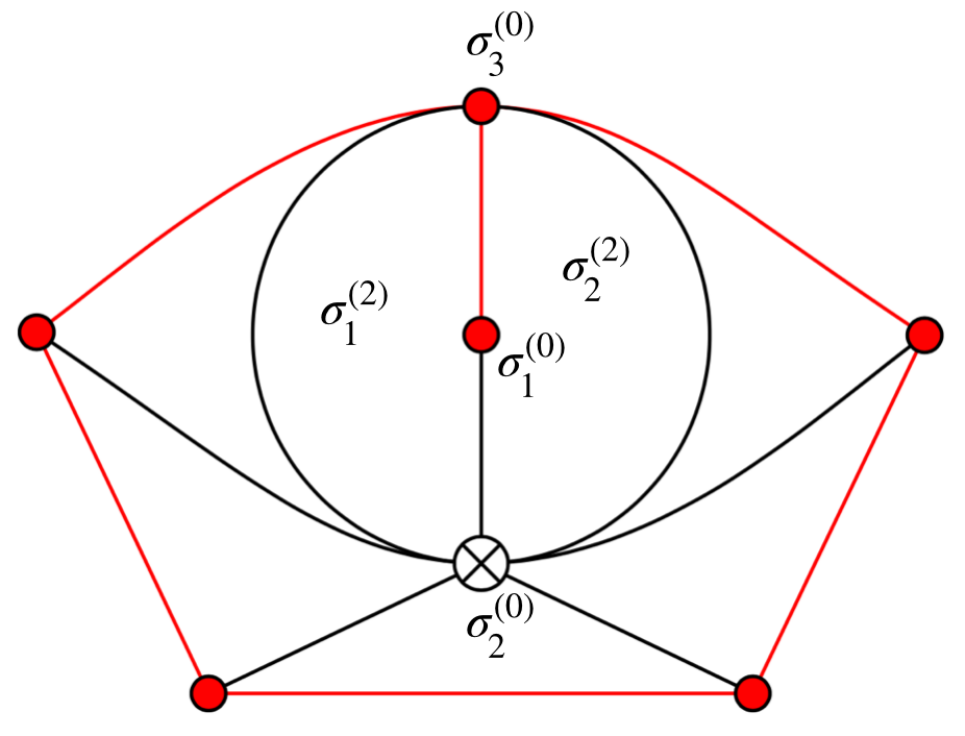}
\caption{The situation of Fig.\ \ref{fig:hat}, enlarged by several additional triangles. The link of the vertex $\sigma^{(0)}_2$ is shown in red, 
illustrating that lk$(\sigma^{(0)}_2)\ncong S^{(1)}$. 
}
\label{fig:flathat}
\end{figure}
It may not be immediately obvious that this is equivalent to our earlier definition in terms of distinct vertices, 
but this can be proven in a straightforward way by considering 
the general neighbourhood of a vertex (see e.g.\ Appendix A of \cite{Niestadt}).
\begin{figure}[t]
\centering
\includegraphics[width=0.7\textwidth]{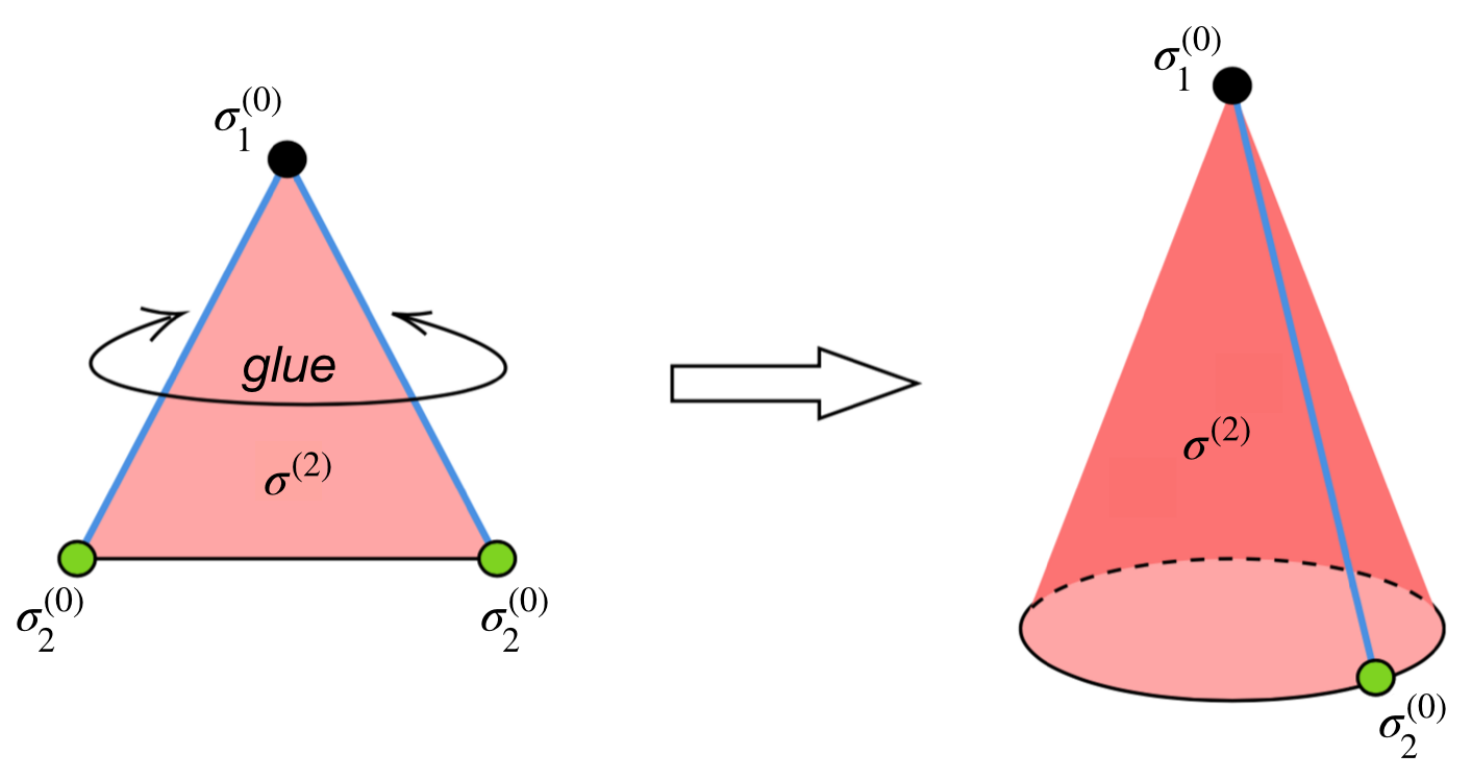}
\caption{The triangle $\sigma^{(2)}$ is glued to itself along two of its sides, and its list of vertices $\{ \sigma^{(0)}_1,\sigma^{(0)}_2,\sigma^{(0)}_2\}$ 
contains one vertex twice. This is only permitted as part of a maximally degenerate triangulation $T\!\in\! {\cal T}_m$, but not for members of the more regular ensembles
${\cal T}_c$ and ${\cal T}_r$. 
}
\label{fig:maxdeghat}
\end{figure}

Moving on to the restricted degenerate ensemble ${\cal T}_r$, the triangles and edges of its configurations are still described by lists of distinct vertices, but these lists 
are no longer required to be unique. As a consequence, two distinct edges can now share the same pair of vertices, thus forming a loop of length 2,
and two distinct triangles can share all of their vertices. An example of the latter is shown in Fig.\ \ref{fig:hat}. Fig.\ \ref{fig:flathat} illustrates that
this local configuration also violates condition (\ref{conds}). It shows the conical hat seen from above as part of a schematic planar diagram.  
We have added some extra triangles that share the vertex $\sigma^{(0)}_2$, to be able to draw a complete link lk$(\sigma^{(0)}_2)$ 
for this vertex. (The argument is not affected by how many triangles we add.) 
The key observation is that this link does not have the form of a one-sphere, because of the ``dangling" edge connecting the vertices $\sigma^{(0)}_1$ and $\sigma^{(0)}_3$, and therefore cannot be part of a combinatorial triangulation. One easily convinces oneself that
the dual $\phi^3$-graph of a restricted degenerate triangulation can contain self-energy, but no tadpole subgraphs. 

\begin{figure}[b]
\centering
\includegraphics[width=0.48\textwidth]{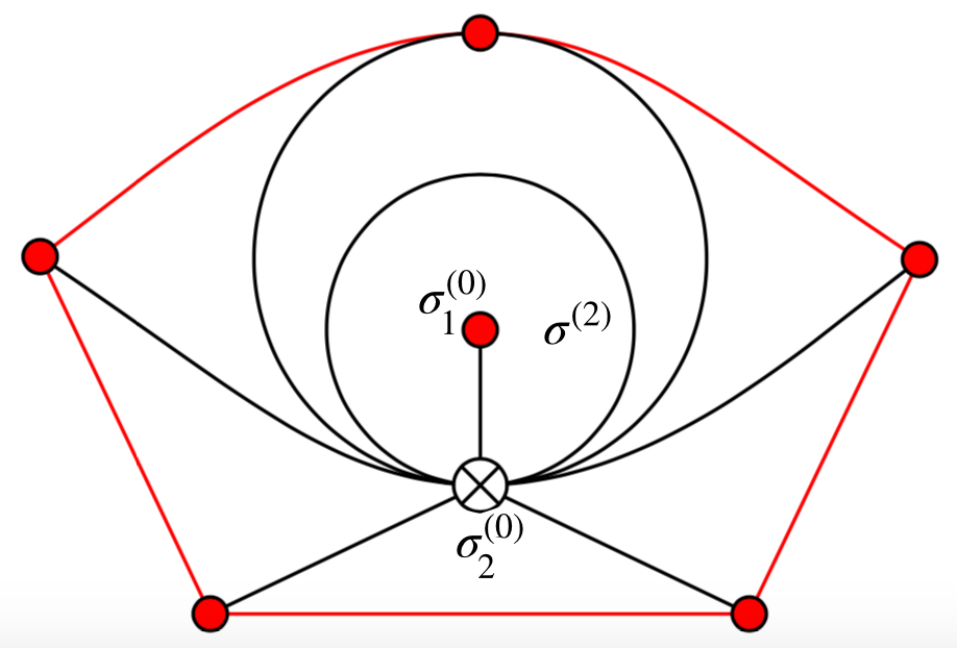}
\caption{The situation of Fig.\ \ref{fig:maxdeghat}, enlarged by several additional triangles. The link of the vertex $\sigma^{(0)}_2$ is shown in red, 
illustrating that also here lk$(\sigma^{(0)}_2)\ncong S^{(1)}$. 
}
\label{fig:maxdegflat}
\end{figure}

Lastly, when defining maximally degenerate triangulations, we also remove the requirement that vertices on a list must be all different. 
In other words, for any $T\!\in\! {\cal T}_m$, the lists of vertices describing its edges and triangles are not necessarily unique, and the vertices within these lists are 
not necessarily distinct. This allows for situations where a single edge forms a closed loop, an example of which is depicted in Fig.\ \ref{fig:maxdeghat}.
For the dual $\phi^3$-graphs it means that not only self-energy, but also tadpole subgraphs are now allowed. 
The local situation shown in Fig.\ \ref{fig:maxdeghat} also violates condition (\ref{conds}), as illustrated by Fig.\ \ref{fig:maxdegflat}. 
Here, we observe that the link lk$(\sigma^{(0)}_2)$ is disconnected,
consisting of a circle and the vertex $\sigma^{(0)}_1$. A general analysis of how the links of zero- and one-simplices violate (\ref{conds}) in configurations 
belonging to the restricted and maximally degenerate ensembles can be found in \cite{Niestadt}.

\end{document}